\documentclass[prd,aps,10pt,twocolumn,floatfix,nofootinbib]{revtex4-1}
\pdfoutput=1
\usepackage{lipsum}
\usepackage{textcomp}
\usepackage{amsmath}
\usepackage{amsfonts}
\usepackage{amssymb}
\usepackage{graphicx}
\usepackage{hyperref}
\usepackage{subfigure}
\usepackage[usenames, dvipsnames]{color}

\begin{document}

\title{A novel scheme for simulating the force-free equations: boundary conditions and the evolution of solutions towards stationarity} 

\author{F. L. Carrasco}
\email{fcarrasc@famaf.unc.edu.ar}
\author{O. A. Reula}
\email{reula@famaf.unc.edu.ar}
\affiliation{ FAMAF-UNC, IFEG-CONICET, Ciudad Universitaria, 5000, C\'o{}rdoba, Argentina.}


\date{\today}

\begin{abstract}
Force-Free Electrodynamics (FFE) describes a particular regime of magnetically dominated relativistic plasmas, 
which arises on several astrophysical scenarios of interest such as pulsars or active galactic nuclei.
%
In this article, we present a full 3D numerical implementation of the FFE evolution around a Kerr black hole. 
The novelty of our approach is three-folded: 
i) We use the ``multi-block'' technique \cite{Leco_1} to represent a domain with $S^2 \times \mathbb{R}^{+}$ topology within a stable finite-differences scheme. 
ii) We employ as evolution equations those arising from a covariant hyperbolization of the FFE system \cite{FFE}. 
iii) We implement stable and constraint-preserving boundary conditions to represent an outer region given by a uniform magnetic field 
aligned or misaligned respect to the symmetry axis.

We find stationary jet solutions which reach equilibrium --through boundary conditions-- with the outer numerical surface.
This is so, even when the outer boundary is located very close to the central region (i.e. $r_{out}\sim 10M $).
These numerical solutions reproduce most of the known results 
for analogue astrophysical settings. 

\end{abstract}


\maketitle

\section{Introduction}

Powerful relativistic jets are often observed in various astrophysical phenomena expanding a very wide range of space-time scales, from pulsars to active galactic nuclei (AGN). 
The mechanism to power such energetic phenomena --even though not jet fully understood-- is believed to reside on the magnetospheres around compact (spinning) objects 
like neutron stars or black holes; where intense magnetic fields, accompanied by charged particles (plasma), tap rotational energy from the central object
and induce strong Poynting fluxes \cite{Lynden1969, goldreich, Blandford}. 
While pulsars admit a classical electrodynamical interpretation in terms of \textit{unipolar induction} \cite{goldreich}, 
the AGN scenario is more subtle, since it involves the question of how to extract energy from a black hole.
The general picture for AGNs was developed in the seminal work of Blandford \& Znajek \cite{Blandford}, 
providing a plausible astrophysical mechanism to extract the BH energy electromagnetically, in a form of generalized Penrose process \cite{lasota2014}.
They demonstrated the existence of stationary electromagnetic fields that possess outgoing Poynting fluxes,
enabled by the presence of a rarefied plasma around a (slowly) spinning BH with a magnetized accretion disk.
This effect is now broadly known as the Blandford-Znajek (BZ) mechanism.

Force-Free Electrodynamics (FFE), describes a particular regime of magnetically dominated plasmas
that plays a key role in the physics of both neutron stars and black hole magnetospheres.
In these regimes, regarded as the vanishing particle inertia limit of ideal relativistic magnetohydrodynamics (MHD) \cite{komissarov2002},
the electromagnetic field obeys a modified --nonlinear-- version of Maxwell equations, 
while the plasma only accommodates as to locally cancel-out the Lorentz force.
Blandford \& Znajek first argued that, under typical astrophysical conditions, 
the magnetosphere around a spinning black hole would be composed of a tenuous plasma arising from a self-regulatory particle production cascade,
so that the dynamics there would be effectively force-free. 
This has been later supported by full MHD numerical simulations (e.g.~\cite{mckinney2004, komissarov2004, mckinney2005, komissarov2005, tchekhovskoy2008, tchekhovskoy2010, tchekhovskoy2011, mckinney2012}), 
suggesting the plasma density is very low away from the disk and especially in the funnel region over the jet.
And although a ``real relativistic jet'' is expected to deviate significantly of the force-free regime at large distances, 
it was argued that the jet power would be determined entirely by the initial force-free zone \cite{tchekhovskoy2010}.

Most MHD simulations also agreed upon the presence of a (self-consistently generated) large-scale magnetic field threading the black hole.
While the strength and topology of this field is not completely clear (since is highly dynamical and dependent on the details of the accretion flow),
it has been suggested \cite{mckinney2005} that near the rotating BH the dominant field geometry would be given by a vertical contribution.
Such magnetic field around the central region is believed to play two important roles: (i) to power the jet by extracting BH rotational energy through the BZ mechanism; 
and (ii) to confine the jet. 

Considerable analytical and numerical efforts have concentrated on the study of black hole magnetospheres under the force-free approximation 
(see e.g. \cite{Macdonald1984, komissarov2001, komissarov2002, Komissarov2004b, komissarov2007, Palenzuela2010Mag, alic2012, ruiz2012, shapiro2013, Gralla2013, Gralla2014, Yang, Pfeiffer2015, gralla2015}).
Numerical simulations started to look at more realistic astrophysical scenarios beyond the monopole, split-monopole and paraboloidal field configurations of the original BZ work. 
In the AGN context, the black hole magnetosphere was usually regarded as embedded on an (asymptotically) uniform vertical magnetic field,  
supported by the currents generated on a distant accretion disk. %
This was referred as the \textit{magnetospheric Wald problem}, in allusion to an exact static electro-vacuum solution constructed by Wald \cite{Wald}.  
It was then noticed that --as for any dipolar type configuration-- the numerical solutions develop an equatorial current sheet within the ergosphere, 
in which the force-free approach breaks down and some sort of electrical resistivity must take over.
And it has been suggested that such current sheet might play a crucial role on determining the magnetosphere structure.
Thus, this plasma-filled version of the Wald problem was signaled as \textit{``an ultimate Rosetta Stone for the research in black hole electrodynamics''} \cite{Komissarov2004b}.


In spite of the fact that the force-free equations have been around for several years, the mathematical details regarding
their initial and boundary value formulation were not fully developed until recently.
Furthermore, it has been shown that a direct formulation of force-free electrodynamics renders a weakly hyperbolic
set of evolution equations \cite{Pfeiffer}, and hence, an ill-posed problem. 
On that article (and on a subsequent work \cite{Pfeiffer2015}) these authors show the system was not only strongly hyperbolic but symmetric hyperbolic,
by presenting suitable reformulations of the theory in a particular $3+1$ decomposition.
Following a different approach, we have constructed in \cite{FFE} a covariant hyperbolization for the FFE system 
relying on Geroch's generalized symmetric hyperbolic formalism \cite{Geroch}. 
Since our construction involves an explicit symmetrizer, it has allowed us to find an appropriate set of evolution equations. 
Moreover, we were able to covariantly extend the system outside the constraint submanifold (namely, the surface of all tensor fields $F_{ab}$ satisfying $F_{ab}^{*}F^{ab} = 0$). 
The extension was built in a way that guarantees the system remains well-posed even outside of the constraint surface, 
a essential property for a numerical implementation, as one knows it is not feasible to enforce the constraint condition exactly.
In the present work, we use --for the first time-- the set of equations derived in \cite{FFE}, to numerically evolve the FFE system.

To implement the problem numerically, we shall consider a computational domain with $S^2 \times \mathbb{R}^{+}$ topology; 
that is, a region 
foliated by successive concentric spherical layers. 
This allows to excise the spacetime singularity from the domain and, on the other hand, to facilitate the implementation of boundary conditions.
To smoothly cover a region with this topology one is forced to employ multiple coordinate patches, and thus, multiple grids.
The challenging aspect of doing so, is in how to ensure a suitable transfer of information among the different grids involved.
We adopt to that end, the \textit{multi-block approach} \cite{Carpenter1994, Carpenter1999, Carpenter2001} (see also \cite{Leco_1} for the present implementation on curved backgrounds), 
which enables the construction of stable finite-differences schemes for computational domains with several grids that only abut\footnote{
Only points at boundaries are common to different grids.}. 
The method relies on the use of finite difference operators satisfying \textit{summation by parts} and \textit{penalty techniques} to transfer information between the grids.
It essentially consist in the addition of penalty terms to the evolution equations at the interfaces among different patches;   
these terms are constructed using the characteristic information of the particular evolution system and guarantees:
(i) a consistent transfer of information between the different grids, and (ii) the derivation of energy estimates at the semi-discrete level used to ensure numerical stability.

We shall picture our computational domain as immersed on an ambient electromagnetic configuration, which in our case, would be the uniform magnetic field of the magnetospheric Wald problem. 
Then, we use the \textit{penalty technique}, combined with such fixed ``external solution'', to set the incoming (characteristic) physical modes at the outer numerical edge.
This way, and together with a method to handle the magnetic divergence-free constraint at the boundary (adapted from \cite{Mari}), 
we achieve stable and constraint-preserving boundary conditions for the astrophysical scenario of interest. 
In contrast, the usual treatment one founds in the literature consist on placing the outer numerical edge far from the central region 
(typically at a radius $\sim 100 M$, or even further away) and setting maximally dissipative conditions there. 
That is, no-incoming modes from the external surface: nothing comes in and any physical signal reaching the boundary would leave the numerical domain.
Therefore, the jet solutions obtained under this setup were usually referred as \textit{quasi-stationary} configurations
(see e.g. \cite{Palenzuela2010Mag,Palenzuela2011}), reflecting the fact that the interior jet solution is not at ``equilibrium'' with the outer boundary. 

The main result of this article may be stated as follows: within our numerical approach, involving the evolution of a new set of FFE equations
and a novel treatment for the boundary conditions, we have obtained truly stationary jet solutions. 
These solutions are consistent with those found in previous works for similar astrophysical settings (like e.g. \cite{Komissarov2004b, Palenzuela2010Mag, Yang}). 
Both the aligned and misaligned cases exhibit a collimated Poynting flux along the direction of the asymptotic magnetic field
and energy extraction through the BZ mechanism; 
a current sheet develops within the ergosphere at the plane normal to the flux, 
and is found to supply significant amounts of EM energy to the total emitted power;
the dependence of the net electromagnetic flux on black hole spin and inclination angle is analyzed as well.
We have emphasized our numerical solutions are stationary since, even locating the outer numerical edge close to the central region, 
the dynamical fields --after the initial transient-- always ``equilibrate'' with the boundary and remains stable for long times. 

This paper is organized as follows: We begin in Sec.~II by introducing our numerical implementation, 
including a brief description of the \textit{multi-block approach} and a detailed discussion of the boundary conditions adopted.
In Sec.~III we present the numerical results on the magnetospheric Wald problem, considering both the aligned and misaligned cases. 
We analyze the influence the numerical boundary (condition and location) has on the results.
Conclusions and some perspectives for future projects are presented on Sec.~IV.
Appendix A review the set of FFE evolution equations (taken from \cite{FFE}) implemented here.
Useful complementary material regarding conservations and the BZ mechanism is provided on Appendix B.
While Appendix C shows further tests to the code, including convergence and an analysis of the constraints behavior.

\section{Numerical Implementation}

We evolve the equations of force-free electrodynamics obtained in \cite{FFE}, 
which we have included again here on an appendix (see equations \eqref{dt_phi},\eqref{dt_E}+\eqref{damp},\eqref{dt_B})
in order to keep this article self-contained.
Our numerical implementation is based on the \textit{multi-block approach} \cite{Leco_1, Carpenter1994, Carpenter1999, Carpenter2001} 
and on a computational infrastructure developed in \cite{Leco_1}, where a particular multiple patch structure has been equipped with the Kerr metric in appropriate coordinates.
In this section, we briefly summarize this approach and provide further details on the treatment given to boundary conditions.
We then discuss the initial and boundary data chosen; and finally, we describe how we deal with the current sheet that develops within the ergoregion. \\

\subsection{General Scheme}

We consider a numerical domain consisting on several grids which just abut (i.e. do not overlap), 
commonly referred to as \textit{multi-block approach}. 
The equations are discretized at each individual subdomain by using difference operators constructed to satisfy summation by
parts\footnote{A property representing the discrete analogue of integration by parts at the continuous level.}  (SBP). 
In particular, we employ difference operators which are third-order accurate at the boundaries and sixth-order on the interior.
\textit{Penalty terms} \cite{Carpenter1994, Carpenter1999, Carpenter2001} are added to the evolution equations at boundary points.
These terms penalize possible mismatches between the different values the characteristic fields take at the interfaces, 
providing a consistent way of communicate information between the different blocks:
essentially, the outgoing characteristic modes of one grid are matched onto the ingoing modes of its neighboring grids.
At each subdomain, it is possible to find a semi-discrete energy defined by both a symmetrizer of the system at the continuum and a discrete scalar product (with respect to which SBP holds). 
The summation by parts property of the operators allows one to obtain an energy estimate, up to outer boundary and interface terms left after SBP. 
The penalties are constructed so that they make a contribution to the energy estimate which cancels inconvenient interface terms, thus providing an energy estimate which covers the whole integration region across grids. 
Such semi-discrete energy estimates --provided an appropriate time integrator is chosen-- guarantee the stability of the numerical scheme \cite{Leco_2}.
A classical fourth order Runge-Kutta method is used for time integration in our code. 

Each subdomain is handed to a separate processor, while the information required for the interfaces treatment is communicated among them by the \textit{message passing interface} (MPI) system.
The computation for each grid may be, as well, parallelized by means of OpenMP.

We incorporate numerical dissipation to the code through the use of adapted Berger-Oliger operators \cite{Tiglio2007}.
We handle the output of our numerical simulations with \textit{VisIt} \cite{VisIt}, 
a visualization tool used to make most of the plots in this article. 

\subsection{Multiple Patch Structure}\label{sec:grids}

\begin{figure}[h!]
  \begin{center}
\includegraphics[scale=0.26]{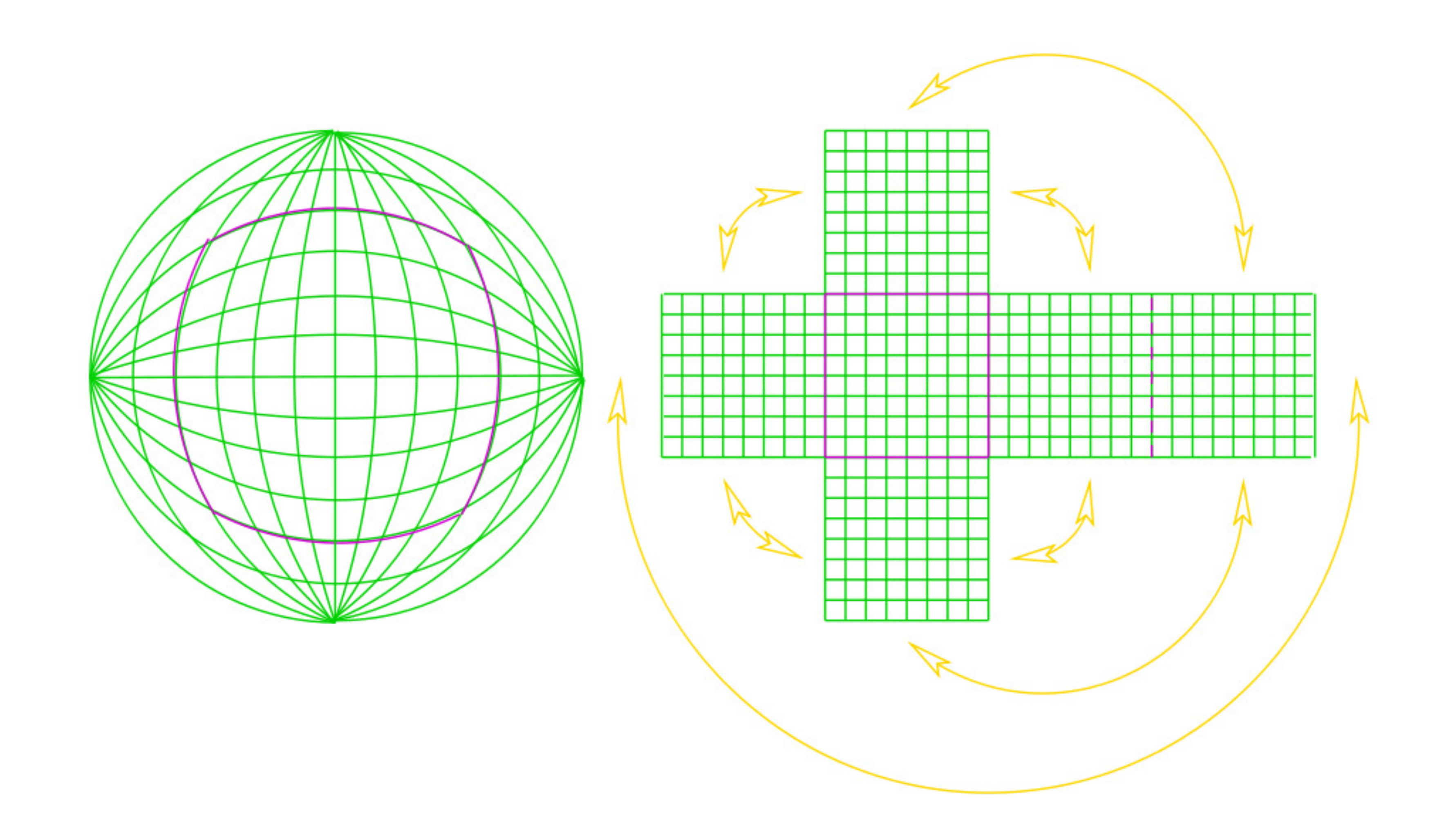}
  \caption{\textit{Cubed sphere coordinates:} 
  six Cartesian patches to cover the sphere, with points only overlapping at boundaries.}
 \label{fig:cubed} 
 \end{center}
\end{figure}

\begin{figure}
  \begin{center}
\includegraphics[scale=0.2]{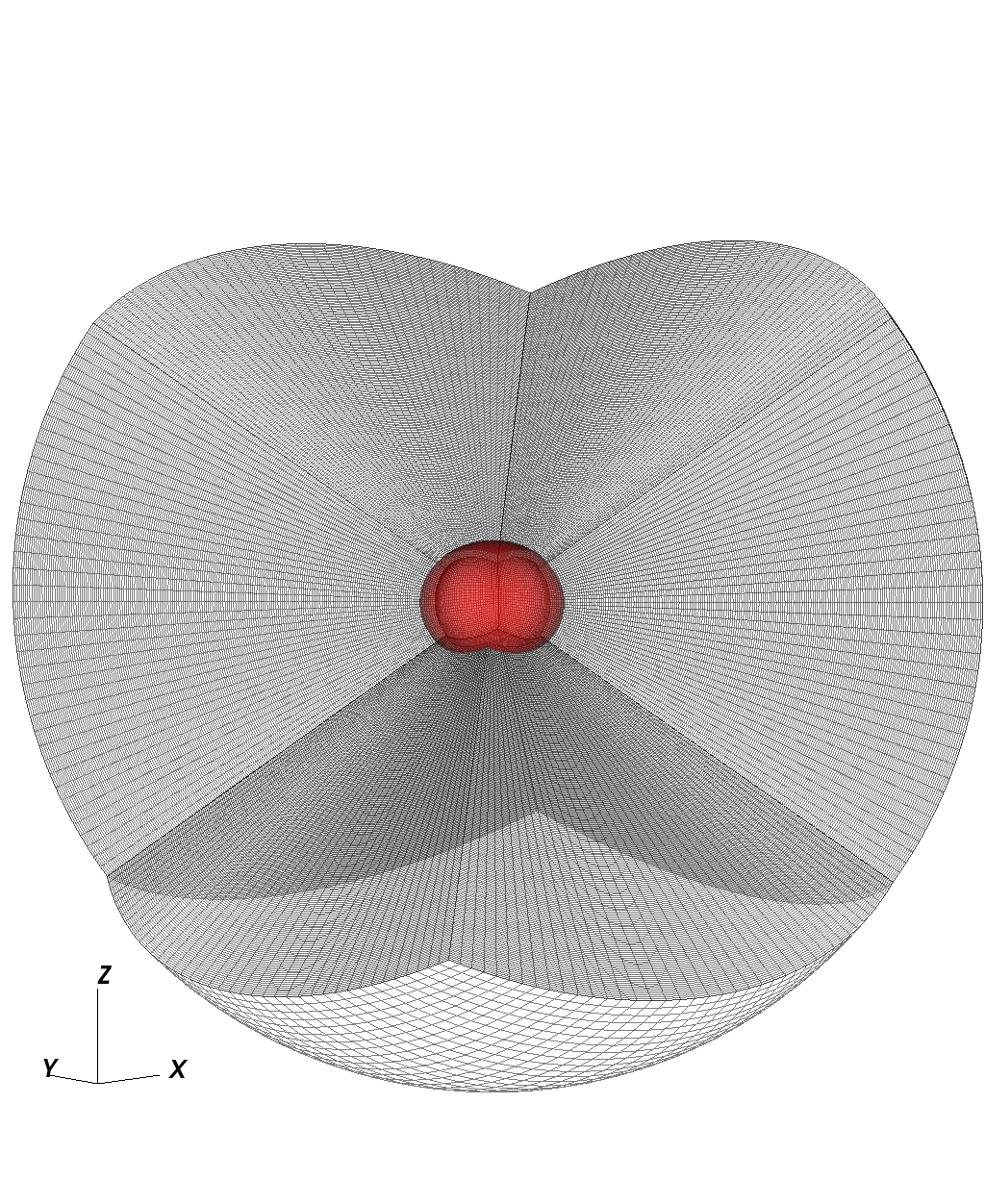}
  \caption{Half of the computational domain for a typical resolution (grid-points are at the intersections of the black lines). 
  Three grids are shown, from the six that makes a spherical shell covering the range between $r=1.35M$ to $r=11.35 M$.
  Further spherical shells (like this one) can be added to cover larger radial distances.
  The semitransparent (red) surface represents the event horizon of a Kerr black hole with $a=0.9$.}
 \label{fig:mesh} 
 \end{center}
\end{figure}

We want to represent a foliation of the Kerr spacetime on a computational domain with the topology $S^2 \times \mathbb{R}^{+}$. 
A relatively simple multi-block structure for $S^2$ is given by the so called \textit{cubed sphere coordinates}, 
represented by a set of six patches (as illustrated on Figure \ref{fig:cubed}).  
The idea is to write the background metric in Kerr-Schild form and use the \textit{cubed sphere coordinates} for the angular directions.
This scheme was already considered in \cite{Leco_1}, where the explicit definitions for this set of coordinates are provided.
In this way, the three-dimensional space can be though as being foliated by successive layers of spherical concentric surfaces.
Our computational domain is thus a spherical shell, like the one depicted on Fig.~\ref{fig:mesh}.
%
The inner edge is located at some radius $r_{in}$, always inside the black hole horizon (for any choice of spin parameter),
and hence, no boundary condition is needed there, as it constitute a purely-outflow surface.
The outer boundary, on the other hand, extends to a radius, $r_{out}$, that will range from $10 M$ to $100 M$ (depending on the case under consideration).  
Notice that $r$ here (and hereafter) denotes the radius in the isotropic coordinates of the Kerr-Schild foliation, rather than the usual Kerr-Schild or Boyer-Lindquist radial coordinates.

The metric in Kerr-Schild form can be written as $g_{ab} = \eta_{ab} + H \ell_{a} \ell_{b}$, 
where $\eta_{ab}$ is the flat metric and $\ell_a$ is certain null co-vector (both with respect to the flat and the whole metric).
For visual representation, throughout this article, we will present our results in the Cartesian coordinates $\{x,y,z\}$ 
associated with the flat part of the metric\footnote{Sometimes referred as the Kerr-Schild ``Cartesian'' coordinates, or the Kerr-Schild frame (see e.g.~\cite{Visser2007}).}. %
That is why the event horizon, the red semitransparent surface in Fig.~\ref{fig:mesh}, does not look spherical. 

The typical resolution employed in the simulations consist on $41\times41$ points for the angular directions at each grid\footnote{
Totalizing a number of around $9600$ points for each of the spherical layers that foliate space.}, 
and $201$ points to cover a radial distance of $10 M$. 
Fig.~\ref{fig:mesh} illustrates this typical resolution to provide the readers with an intuitive feel of grid-points density.
In some cases we have doubled the resolution (i.e. $81\times81\times401$ for each grid), as in Figs.~\ref{fig:PF_slice_aligned} and \ref{fig:PF_slice_misaligned};
and some calculations were done using a coarser resolution (half of the typical one) as those shown in Figs.~\ref{fig:PF_spin_dep} and \ref{fig:PF_ang_dep}.

\subsection{Boundary Conditions}

We must prescribe boundary conditions (only) on the external surface of our numerical domain, i.e. the sphere of radius $r_{out}$. 
That is, we need to set --at each point over this surface-- the incoming characteristic modes associated with our particular evolution system. 
There are in general two kind of conditions: those corresponding to the physical scenario one wishes to represent and those related with the preservation of the constraints.
We shall rely on the penalty method to enforce the physical modes, while we will use alternative methods to ensure no constraint violations arise from the boundary.

For the evolution equations we want to implement, there are four physical modes 
(the fast magneto-sonic and Alfven waves) and three (unphysical) constraint modes.
Since the theory is nonlinear, these different subspaces might degenerate at some points during the evolution, 
and thus, a careful analysis for the characteristic structure of the system is needed.
This was already done by the authors in Appendix A of reference \cite{FFE}, where all the possible degeneracy cases were examined in detail. 
With this information at hand we are now in conditions to appropriately modify the evolution equations at the outer boundary points.

\subsubsection{Physical condition}

As previously discussed, the idea here is to use the \textit{penalty terms}, which may be written:
\begin{equation}
  \partial_t U^{\alpha} \rightarrow \partial_t U^{\alpha} + \sum_{i} \left( \frac{\lambda_i}{\sigma_{oo} \Delta x } \right) P^{\alpha}_{(i) \beta} \left( U^{\beta}_{ext} - U^{\beta} \right) \label{penalty}
\end{equation}\\
with $U^{\alpha} :=  \{ \phi ,  E^i , B^i  \} $ being the set of dynamical fields. 
The corrective terms are given by a sum over all the incoming physical modes 
(i.e. physical characteristic subspaces ``$i$'' with positive eigenvalues, $\lambda_i > 0$), where $P^{\alpha}_{(i) \beta}$ is the projection onto the $i$-subspace, 
and $U^{\alpha}_{ext}$ represents the ``exterior'' solution we want to impose \footnote{This idea of analytically fixing the incoming characteristic modes (``exterior solution'') 
was also used in \cite{Pfeiffer2015} to study the stability of the exact \textit{null$^{+}$} FFE solutions on a Schwarzschild background.}. 
The factors $\sigma_{oo}$ and $\Delta x$ in the expression reflect the dependence on the discrete inner product used and on the grid resolution, respectively. 

\subsubsection{Constraints}

To restrict the entrance of possible violations of the divergence-free constraint ($\nabla \cdot \vec{B} = 0$), 
we adapt to our problem a method proposed in \cite{Mari}. 
The idea is to study the subsidiary system describing the dynamics of the constraint, and then impose the no-incoming condition on these modes.
Such subsidiary system is obtained from the definitions $D:=\mathcal{D}_j B^j$ and $\delta_i := \partial_i \phi$ as,
\begin{eqnarray}
 && \partial_t D = \beta^k \partial_k D +  D~\mathcal{D}_j \beta^j  -  \alpha~\mathcal{D}_j  \delta^j  - \delta^k \partial_k \alpha + \mathcal{D}_j Z^j  \nonumber\\
 && \partial_t \delta^i = \partial^i (\beta^k \delta_k) - \alpha (\partial^i D + \kappa \delta^i ) - (D + \kappa \phi ) \partial^i \alpha - \partial^i W \nonumber
\end{eqnarray}
where $ Z^i \equiv \frac{\alpha}{\tilde{F}} \left[ \hat{\epsilon}^{ijk} r_j \tilde{B}_k + \frac{\tilde{E}^i}{\tilde{B}^2}\tilde{S}^k r_k \right] $ and $W \equiv \frac{\alpha}{\tilde{F}} \tilde{E}^k r_k$.

Its characteristic problem --after some manipulations-- reduces to,
\begin{eqnarray}
 && (\lambda - \beta_m ) \hat{D} =  -\alpha \hat{\delta}_m \nonumber\\
 && (\lambda - \beta_m ) \hat{\delta}^i = -\alpha \hat{D} m^i \nonumber
\end{eqnarray}
where $m^i$ is the unit normal to the boundary, and we have denoted contractions by, $\beta_m \equiv \beta_i m^i $ and $\hat{\delta}_m \equiv \hat{\delta}_i m^i $. 
Hence, the no-incoming condition reads:
\begin{equation}
 \frac{1}{2} (\delta_m - D) = 0 \label{no-incoming} 
\end{equation}

Solving for $D$ and $\delta_m$ from the general system and imposing the condition \eqref{no-incoming},
we finally get the corrective terms at boundary points:
\begin{eqnarray}
 \partial_t \phi  &\rightarrow& \partial_t \phi  + \frac{1}{2} \alpha \left[ \mathcal{D}_j  B^j - m^j \partial_j \phi \right]  \label{normal_phi}\\
\partial_t B^i &\rightarrow& \partial_t B^i  - \frac{1}{2} \alpha m^i \left[ \mathcal{D}_j B^j - m^j \partial_j \phi \right] \label{normal_B}
\end{eqnarray}\\ 

Regarding the algebraic constraint, $G=0$, we observe the evolution equations naturally determines $\partial_t G \sim 0$
at the boundary, whenever the value of the constraint remains small (i.e. $G\sim 0$) at the interior. 
Thus, provided we manage to keep this constraint under control through the evolution, there would be no need to further modify 
the equations at the outer boundary. 

We monitor the behavior of both constraints 
during our simulations to ensure there are no violations entering from the boundary and, 
moreover, that no significant deviations develop within the whole numerical domain.
To see these results we refer the reader to Section \ref{sec:constraints_preservation}.

\subsection{Initial/Boundary Data}\label{sec:data}

We consider a spinning black hole surrounded by a magnetized accretion disk. 
Assuming the disk to be sufficiently distant, the magnetic field configuration it gives rise to would look essentially
uniform within our computational domain. 
The direction of this magnetic field is perpendicular to the disk, and is not necessarily aligned with the symmetry axis of the spacetime. 
We shall then picture our black hole as initially immersed on a uniform magnetic field, aligned or misaligned respect to the rotational axis. 
In isotropic Kerr-Schild coordinates thus reads,
\begin{equation}
 B^{x} = \frac{B_o}{\sqrt{h}} \sin{(\alpha_o)}  \text{,} \quad  B^{y} = 0  \text{,} \quad  B^{z} = \frac{B_o}{\sqrt{h}} \cos{(\alpha_o)} 
 \label{initial}
\end{equation}
The interior solution will be modified during the evolution due to the presence of the plasma around the BH horizon, 
while the exterior region should remain dominated by the uniform magnetic field configuration. 
Thus, we will set this configuration as the ``exterior'' solution ($U^{\alpha}_{ext}$) on equation \eqref{penalty}, as discussed above.  

We want to chose astrophysically relevant values 
for an scenario with a super-massive black hole of mass $M \sim 10^{6-10} M_{\odot}$ 
and a magnetic field of around $B \sim 10^{1-4} G$  \cite{blandford1992}. 
In particular, we adopt a black hole mass $M = 10^{8} M_{\odot}$ and a magnetic field strength $B_o = 10^{4} G$ for later comparison with \cite{Palenzuela2010Mag}.
In the geometrized units of the code (where the mass of the black hole has been set to unity),
and according to the Lorentz-Heaviside units employed for the evolution equations, the magnetic field strength must be then $B_o [1/M] = 1.2 \times 10^{-8}$. \\

\subsection{Current Sheet Treatment} \label{sec:current_sheet}

It is well known that current sheets may develop on black hole magnetospheres.
In particular, dipolar magnetospheric configurations leads to the formation of a strong current sheet at the equatorial plane.
In these regions, the force-free condition $B^2 - E^2 > 0 $ is no longer satisfied and the theory breaks down, physically and mathematically. 
The perfect conductivity approximation fails and a model of electrical resistivity would be required.
Komissarov \cite{Komissarov2004b} analyzed a model of radiative resistivity based on the inverse Compton scattering of background photons 
and concluded that the cross-field conductivity inside the current sheet has to be governed by a self-regulatory mechanism ensuring marginal 
screening of the electric field. Thus, the electromagnetic field at the current sheet is expected to satisfy,
\begin{equation}
 B^2 - E^2 \approx 0 \nonumber
\end{equation}

A simple way of implement this resistivity numerically (even though not very appealing from the mathematical point of view) 
is by reducing the electric field whenever it gets too close in magnitude to the magnetic one. 
This is applied at each iteration of the Runge-Kutta time step. 
In this way, one is effectively dissipating electric field at the current sheet and driving the electromagnetic field into a state very close to $B^2 - E^2 = 0$, as physically expected. 
We do it by follow a similar prescription to the one employed in \cite{Palenzuela2010Mag}, but ``cutting'' the field on a slightly smoother manner:  
\begin{equation}
 E^i \rightarrow f\left( \frac{|E|}{|B|}\right) E^i \nonumber
\end{equation}
where $f(x)$ is a smooth piecewise function, which equals one for $x \leq 1-2\varepsilon$; is given by a fifth degree polynomial for the interval $ 1-2\varepsilon < x < 1-\varepsilon$;
and is $\frac{1}{x}$ for larger values of $x$. With $\varepsilon$ being a small parameter, generally set to $\varepsilon=0.05$ in the code.


\section{Numerical Results}\label{sec:results}

As discussed in Section \ref{sec:data}, we chose initial/boundary data to study the \textit{magnetospheric Wald problem},
in which a uniform magnetic field dominates the far field region. 
Our numerical solutions goes through an initial dynamical transient where the magnetic field lines twist around and an electric field is induced; 
after which, all the simulations reach a steady state. 
The time to reach such final configurations depends on the location of the outer numerical boundary, to which they equilibrate with.

\begin{figure}[t!]
  \begin{center}
  \begin{minipage}{8cm}
 \subfigure{\includegraphics[scale=0.21]{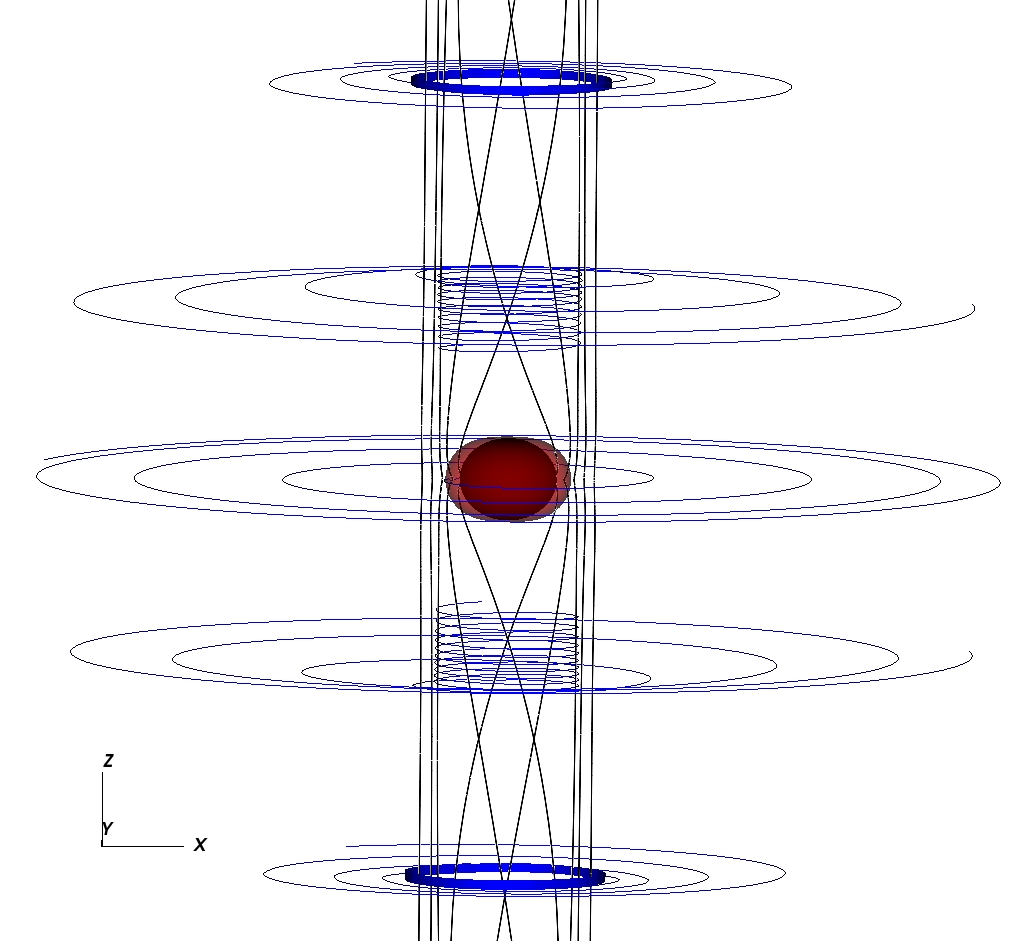}} 
\end{minipage}
\begin{minipage}{8cm}
\subfigure{\includegraphics[scale=0.26]{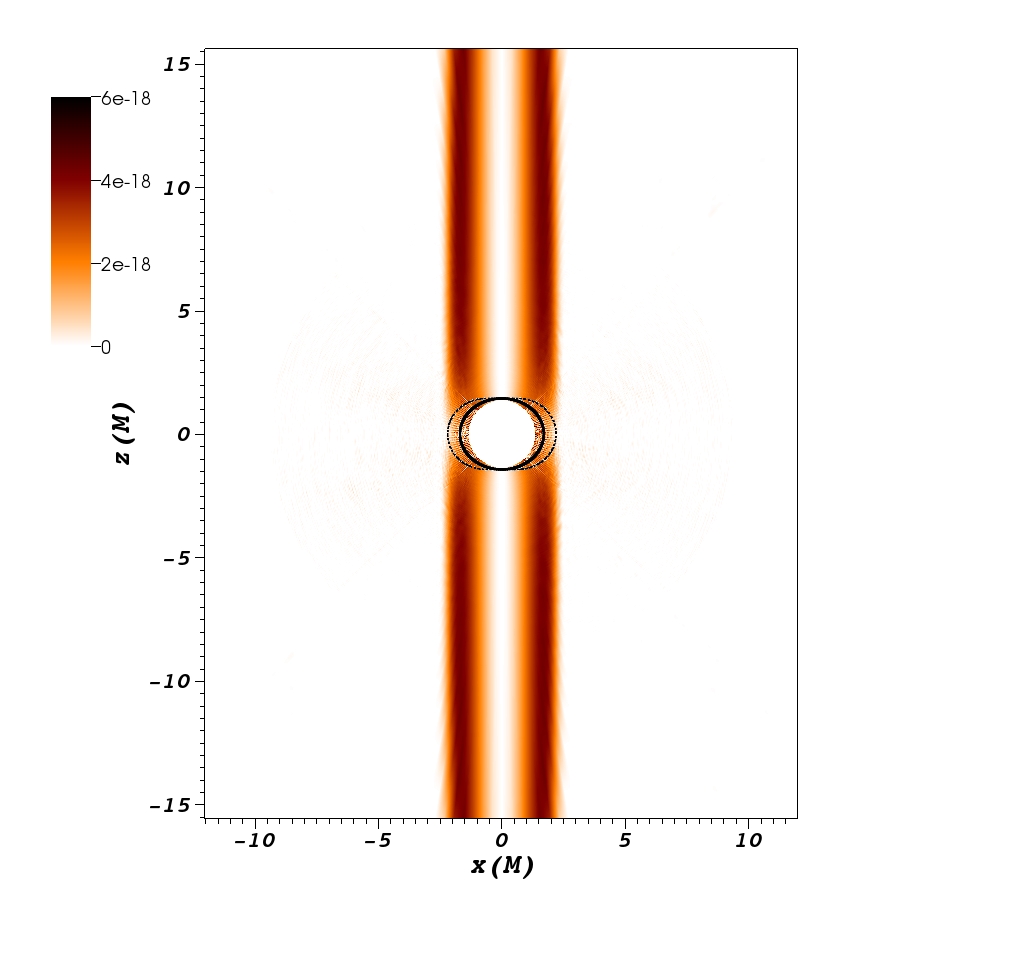}} 
\end{minipage}
  \caption{Aligned case: late time numerical solution ($t=120M$) for a black hole with $a=0.9$. 
   \textbf{Left:} Representative streamlines of the magnetic field (black thick lines) and the electric field (blue lines).
   The solid black surface is the BH horizon, whereas the semitransparent (red) one represents the ergosphere.
   \textbf{Right:} The radial Poynting flux density is shown in color scale at the $x-z$ plane. 
   Thick solid and dotted black lines represents the black hole horizon and ergosphere, respectively.
   }
 \label{fig:PF_slice_aligned} 
 \end{center}
\end{figure}

The late time solutions attained are truly stationary and exhibit collimated flows of electromagnetic energy, as those seen 
on the right panel of figures \ref{fig:PF_slice_aligned} and \ref{fig:PF_slice_misaligned}.
The net flux over any spherical surface, $\Phi(R)$, is always positive, meaning energy is being extracted from the black hole. 
This energy is carried to the asymptotic region in the form of a collimated Poynting flux: the jet. 
\onecolumngrid

\twocolumngrid

We find no significant differences in the late time numerical solutions for initial/boundary data corresponding to the uniform magnetic field 
we use here (also used in \cite{Palenzuela2010Mag}) and the Wald configuration with zero electric field (as considered in e.g. \cite{Komissarov2004b}).
The reason for that, can be easily understood by noticing that the magnetic fields on these two configurations only (slightly) differ near the black hole horizon;
and hence, the boundary data --which is ultimately what determines the final state-- is essentially the same in both cases.

We shall consider first the case where the asymptotic magnetic field is aligned with the symmetry-axis of the black hole.
This will serve to explore some of the known features of these jet solutions: the operation of the Blandford-Znajek mechanism, 
the dependence of this mechanism on black hole spin, the presence of an electronic circuit, and the development of an equatorial current sheet and its effects on the total emitted power.
We will also use this scenario to analyze the influence of the conditions and location adopted for the outer numerical boundary.  
Later, we shall abandon axial symmetry to study the more general case when the asymptotic field is not aligned with the rotation axis of the black hole. 
This way, we will exploit the full potential of our three dimensional code, and confirm some known --though more scarce-- results for this setting.

\subsection{Aligned Case}

In Fig.~\ref{fig:PF_slice_aligned} we have considered a representative (high resolution) numerical solution 
for the aligned case, with black hole spin $a=0.9$.
The general structure of the electric and magnetic fields is depicted on the left panel image,
where it can be seen that all the magnetic field lines which penetrate the ergosphere acquire a toroidal component.
The electric field is predominantly toroidal everywhere, with an induced $z$-component along the jet.
The right panel of Fig.~\ref{fig:PF_slice_aligned} shows the radial\footnote{Here, 
we refer again to a radius $r$ in the isotropic coordinates of the Kerr-Schild foliation, rather than the usual radial coordinate. 
We remark, however, that both densities should look very similar. } 
EM energy flux density, $p^r$ (see \eqref{flux-density}), on the $x-z$ plane. 
It illustrates the highly collimated Poynting flux generated.


A first question one may ask is, whether the energy extraction taking place in these configurations are Penrose-like processes, or not.
In the aligned case, we know the late times solutions are stationary and axi-symmetric.
Thus, we just need to check if our solutions satisfy the Blandford-Znajek condition \cite{Blandford, lasota2014}, 
\begin{equation}
 0 < \Omega_F < \Omega_H \label{BZ-condition}
\end{equation}
within the ergoregion. Where,
\begin{equation}
 \Omega_F := \frac{F_{t \theta}}{F_{\theta \varphi}} \label{frec-rot}
\end{equation}
is the \textit{rotation frequency of the electromagnetic field}, which captures the notion of ``angular velocity of the magnetic field lines'' 
for the stationary and axi-symmetric case (see e.g. \cite{Blandford, Komissarov2004b, Palenzuela2010Mag});
and $\Omega_H := \frac{a}{2M r_H}$, is the frame dragging orbital frequency at the black hole horizon. 

\begin{figure}
  \begin{center}
\begin{minipage}{3.1cm}
 \subfigure{\includegraphics[scale=0.11]{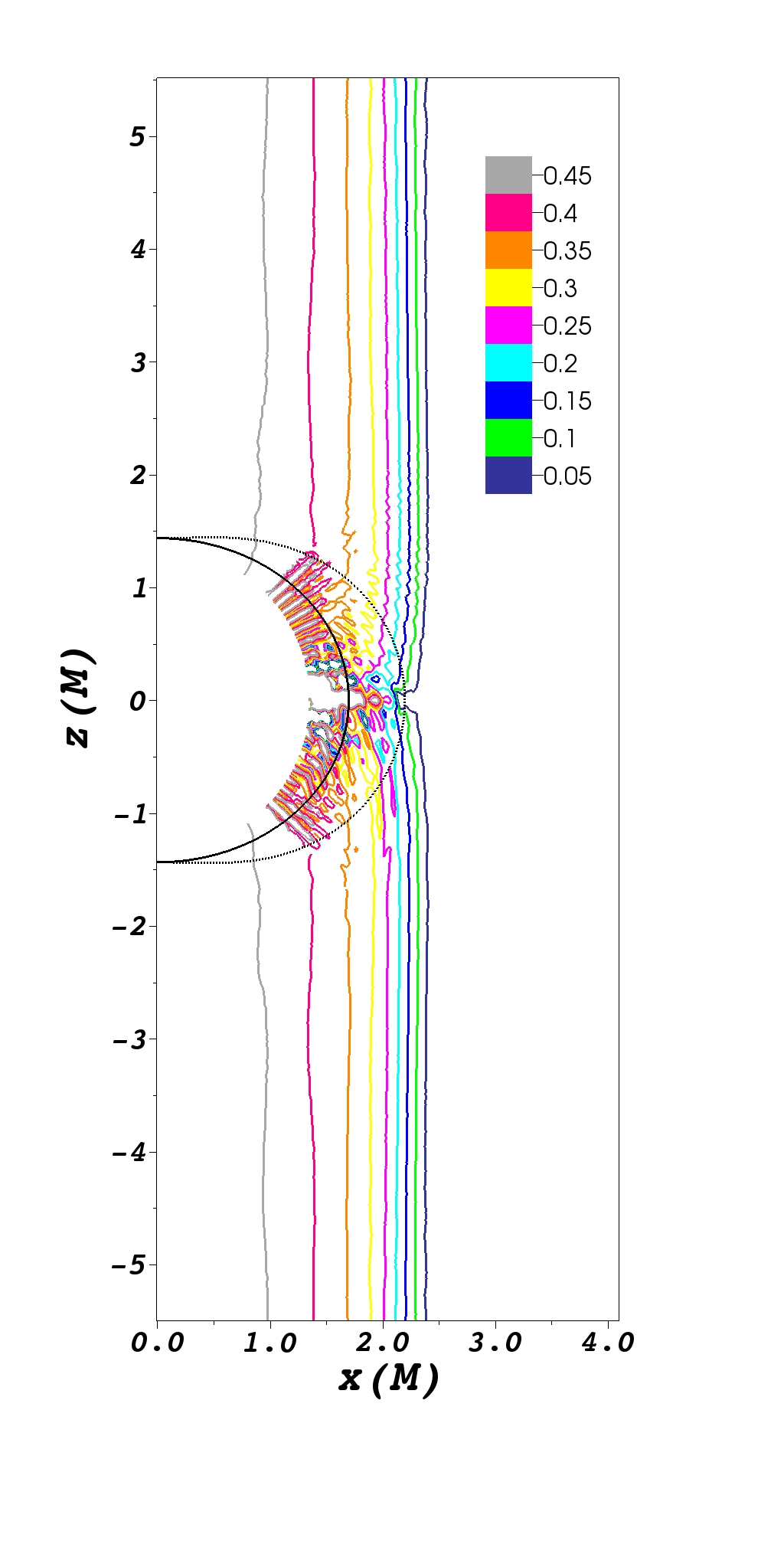}} 
\end{minipage}
\begin{minipage}{5.3cm}
\subfigure{\includegraphics[scale=0.155]{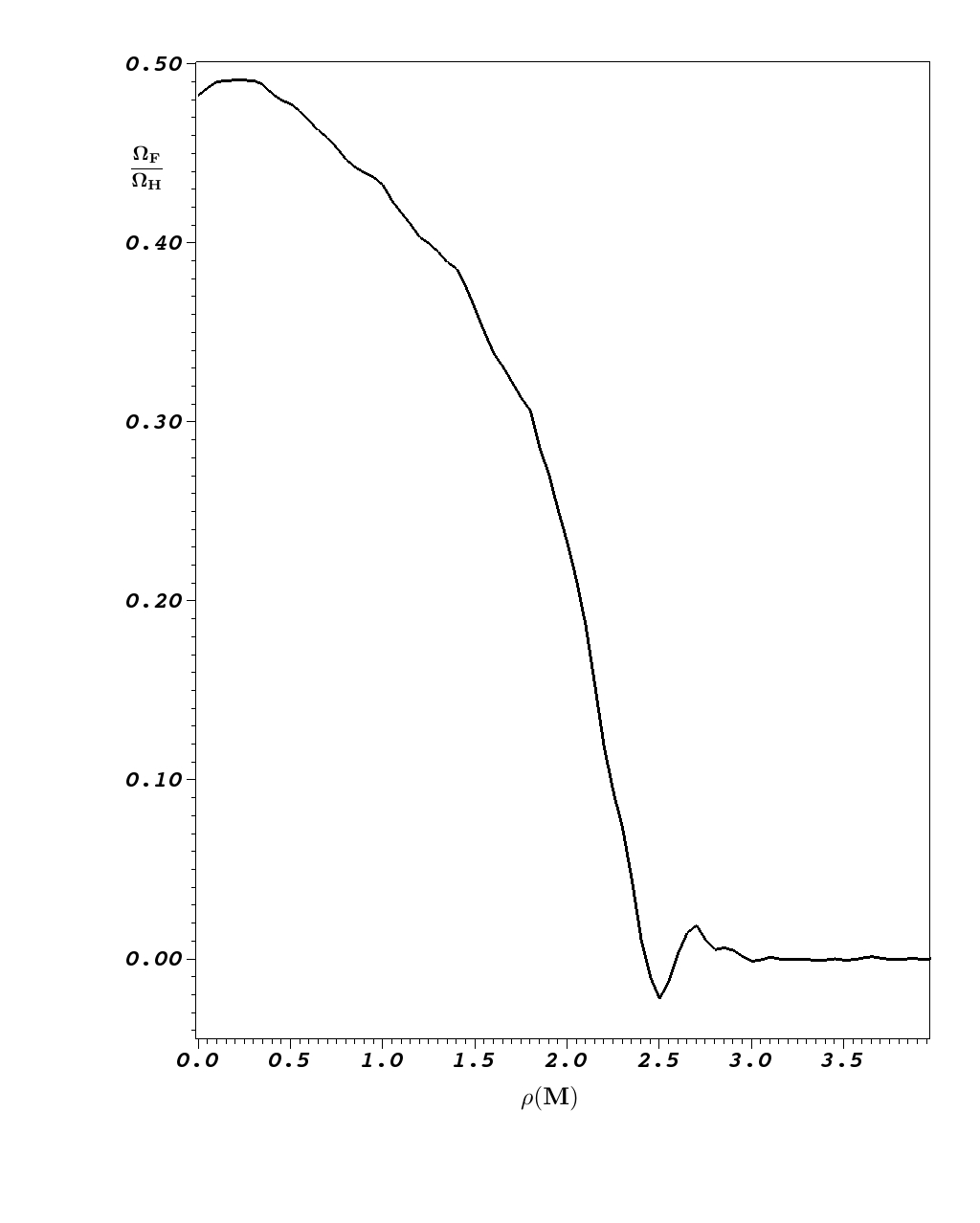}}
\end{minipage}
  \caption{Angular frequency of magnetic field lines at late time ($t=120M$) solution, for a spinning black hole with $a=0.9$. 
  \textbf{Left:} Contour lines of $\Omega_F / \Omega_H $ between $0.05$ and $0.45$ (near the axis) at the $x-z$ plane. 
  Thick solid and dotted lines represents the black hole horizon and ergosphere, respectively. 
  \textbf{Right:} Profile of $\Omega_F / \Omega_H $ across the jet: it shows the distribution on the cylindrical distance from the axis, $\rho$, at $z=4M$. }
 \label{fig:BZ} 
 \end{center}
\end{figure}
In Fig.~\ref{fig:BZ}, we plot the quotient $\Omega_F / \Omega_H $ of a representative late time configuration, 
which confirms that the BZ condition is indeed satisfied. 
The left panel presents contour lines of this quantity on the $x-z$ plane, showing those magnetic field lines crossing the ergoregion acquire angular velocity and fulfill condition \eqref{BZ-condition}.  
Moreover, it can be seen the values attained at the jet region, $\Omega_F \sim 0.5 ~ \Omega_H$, represents a maximum in the power expression, i.e. \eqref{BZ-cond}.
The right panel of Fig.~\ref{fig:BZ}, on the other hand, displays a the distribution of $\Omega_F / \Omega_H $ at $z=4M$. 
It provides with a representative profile (at any height $z$) of the magnetic field lines angular velocity distribution across the jet. 

\subsubsection{Dependence on black hole spin} \label{sec:spin_dep}

\begin{figure}[h!]
  \begin{center}
\includegraphics[scale=0.32]{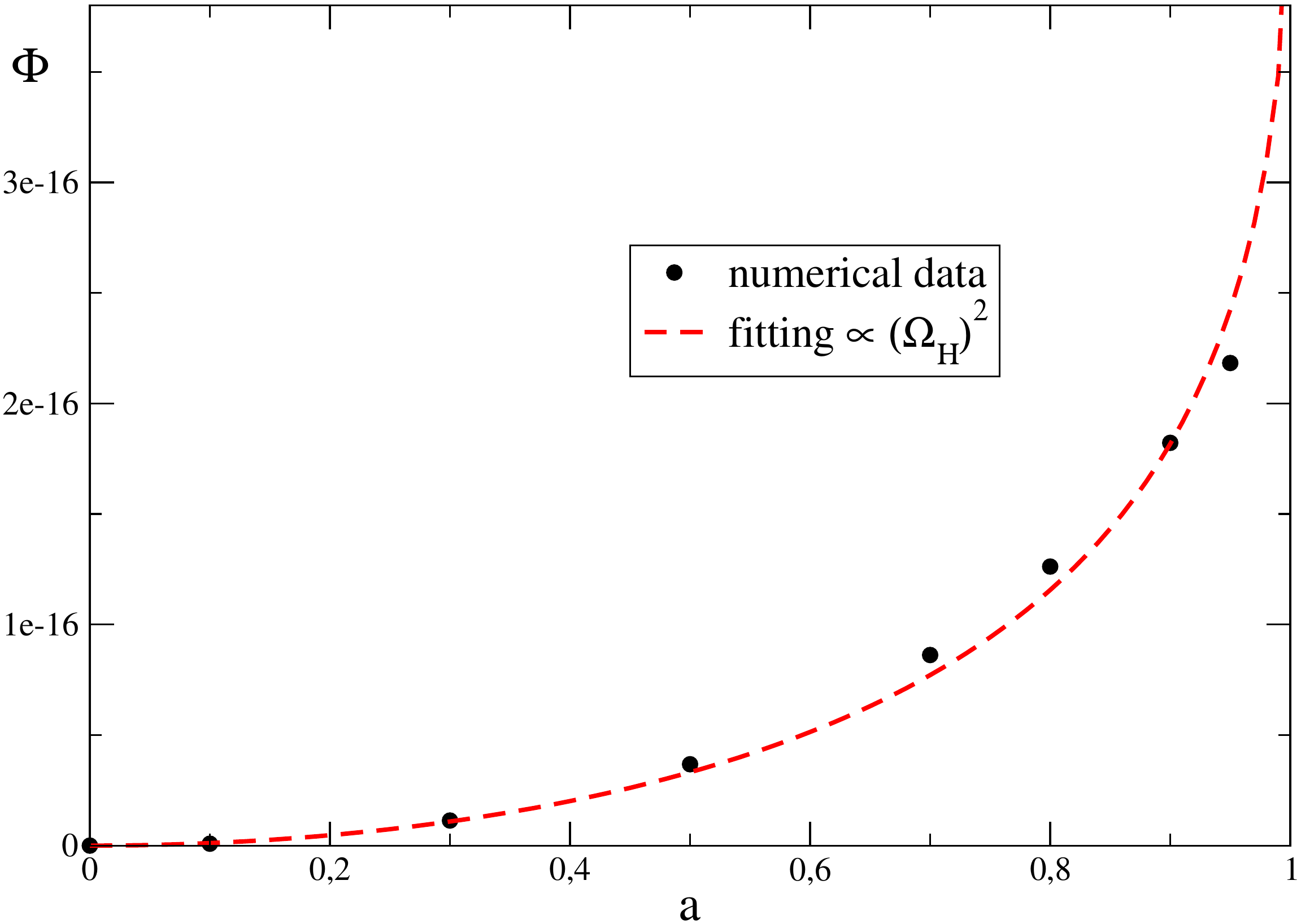}
  \caption{Dependence on black hole spin of the total EM energy flux (in the aligned case). 
  A black hole of mass $M = 10^{8} M_{\odot}$ and a magnetic field of strength $B_o = 10^{4} G$ are considered, for comparison with \cite{Palenzuela2010Mag}.
  The dots corresponds to the net flux (integrated at $r=2.5M$) in our numerical solutions; 
  while the dashed (red) curve is a fit of the form $\Phi \propto \Omega_{H}^2$.}
 \label{fig:PF_spin_dep} 
 \end{center}
\end{figure}

We consider the dependence of the net flux of energy emerging from the black hole on its spin parameter.
It was argued that the total electromagnetic energy flux behaves as $\Phi \propto \Omega_{H}^2$, at first leading order \cite{tchekhovskoy2010, Palenzuela2010Mag}. 
In Fig.~\ref{fig:PF_spin_dep}, we plot $\Phi$ for different values of the spin parameter, together with the curve $\Omega_{H}^{2} (a)$.
It can be seen from the image, that the curve fits the numerical values very well.
We also note these results are in good agreement with those found in fig.~4 of \cite{Palenzuela2010Mag}.
The only significant difference we report, is a factor of (almost) two on the emitted power: 
our net energy flux is roughly twice the one obtained there\footnote{
Recall we chose the strength of the asymptotic magnetic field and black hole mass for comparison with this reference.}.
%
%

\subsubsection{Electronic circuit}

\begin{figure}
  \begin{center}
\includegraphics[scale=0.27]{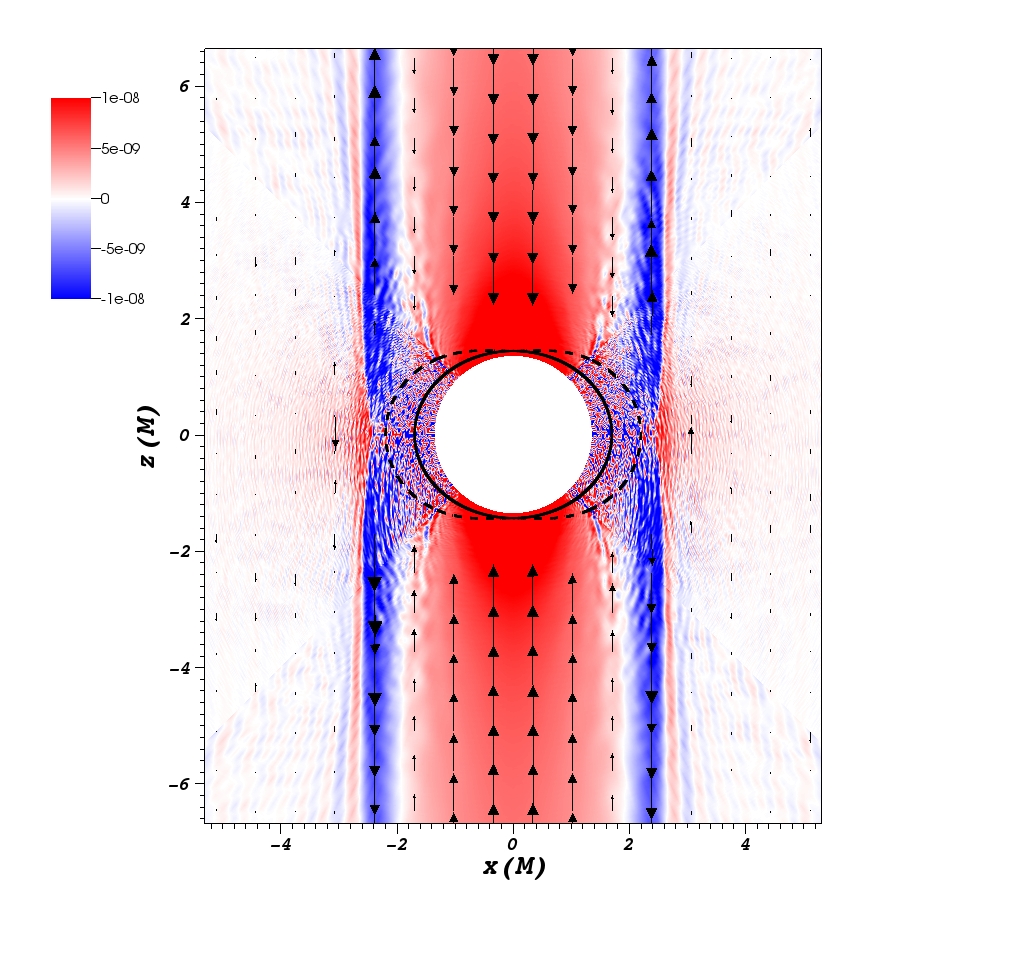}
  \caption{Electronic circuit and induced charge distribution for a late time solution ($t=120M$) on a black hole with $a=0.9$.
  Charge density (color scale) and poloidal electric currents (arrows) are displayed on the $x-z$ plane. 
  Solid/dashed lines represents the black hole horizon and ergosphere, respectively. }
 \label{fig:circuit} 
 \end{center}
\end{figure}

It is interesting to note one can recover part of the information regarding the plasma through Maxwell equation, $j^a \equiv \nabla_b F^{ab}$. 
On Fig.~\ref{fig:circuit}, we display the electric charge density at the $x-z$ plane, along with the induced electric currents found in our late time numerical solutions.
As seen in the image, the current flow in the direction of the symmetry axis into the black hole at the poles and then back out (in the opposite sense) within a cylindrical shell   
starting near the intersection of the ergosphere and the equatorial plane.
Such induced electronic circuit on the magnetospheric plasma is consistent with known qualitative and numerical studies 
(see e.g. \cite{Komissarov2004b, Palenzuela2010Mag});
and is what determines the critical difference with the electro-vacuum case (i.e. no plasma), allowing for the Poynting flux and energy extraction from the black hole.

The ripples observed on the charge density distribution on Fig.~\ref{fig:circuit} are due to the numerical prescription we use to handle the current sheet,
that generate numerical disturbances which then propagate around and dissipate. 
Since computing the charge density involves spatial derivatives, it is particularly sensitive to this numerical noise. 

\subsubsection{Current sheet}

\begin{figure}[h!]
  \begin{center}
\includegraphics[scale=0.25]{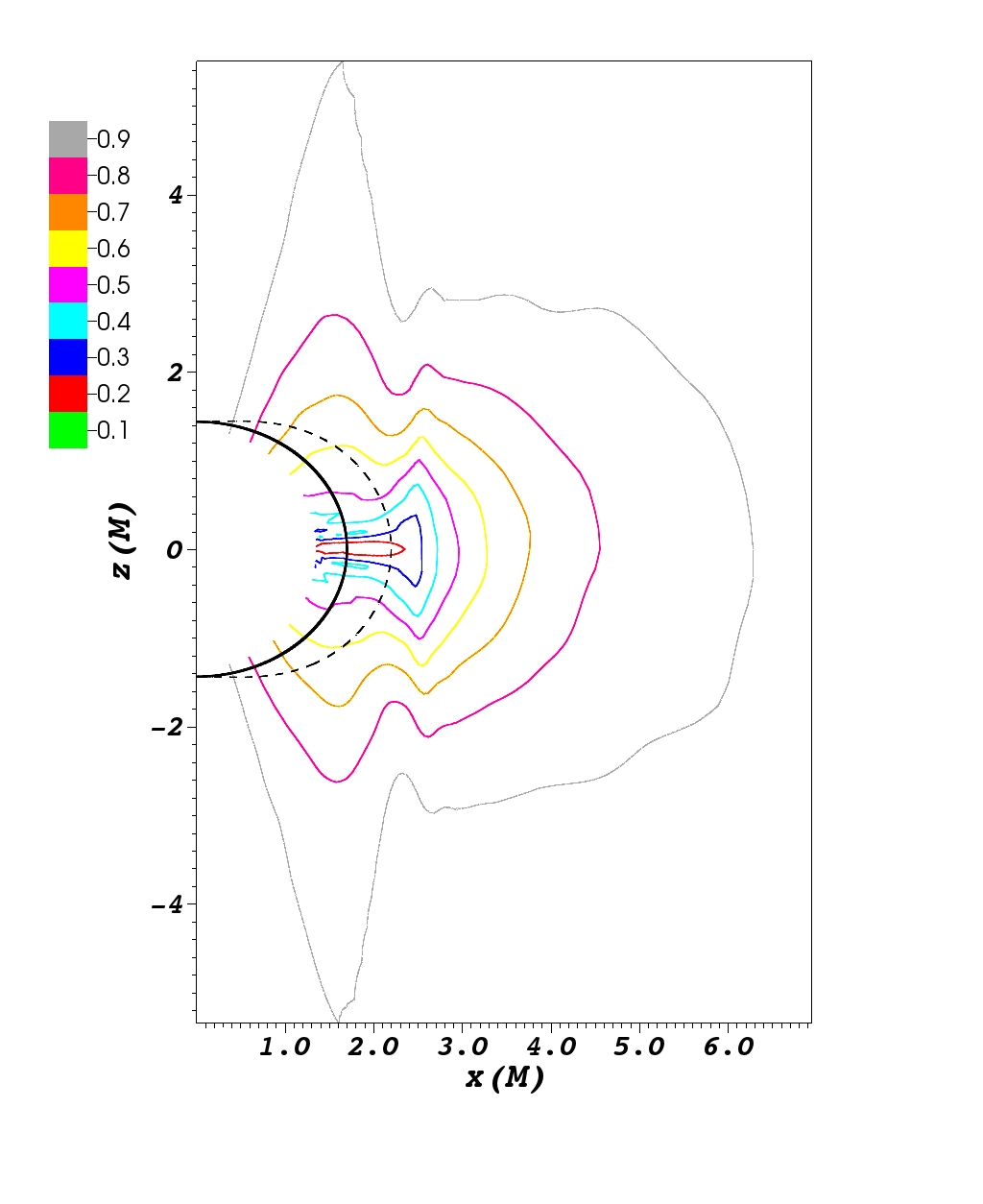} 
  \caption{Contour lines of $\frac{B^2 - E^2}{B^2}$ for a representative late time solution ($a=0.9$) at the $x-z$ plane. 
  It illustrates the structure of the current sheet developed. 
  Solid and dashed black lines represents the BH horizon and ergosphere, respectively.}
 \label{fig:C3} 
 \end{center}
\end{figure}

Since all magnetic streamlines penetrating the ergosphere are ``forced to co-rotate'' with the black hole,
a discontinuity on the toroidal component of the magnetic field is then generated along the equatorial plane (within the ergosphere) and a current sheet develops.
In Fig.~\ref{fig:C3}, we display contour lines of $\frac{B^2 - E^2}{B^2}$ on the $x-z$ plane, 
raging from around $0.1$ at the inner region (close to the equatorial plane) up to $0.9$.
This distribution of $\frac{B^2 - E^2}{B^2}$ illustrates the structure of the current sheet, as it is known that a violation of the magnetic-domination condition 
is taken as evidence that the plasma would have a non-negligible back reaction on the electric field.
At this region, we know our numerical implementation is effectively dissipating electric field (Section \ref{sec:current_sheet}).
We have attempted to restrict the dissipating mechanism to make it operate only inside the black hole horizon\footnote{
By gradually reducing it after the initial transient.}, but failed. 
This fact relates with the observation made in \cite{Komissarov2004b} that the ``gravitationally induced'' electric field cannot be completely screened inside the ergosphere. 
So it seems there is no way to prevent a strong current sheet from developing inside the ergoregion,
and thus, the equilibrium reached by the late time solutions will have this mechanism actively operating there.

Komissarov has suggested that the anisotropic resistivity seems to play a key role in shaping the resulting magnetic field structure
(within the ergosphere at least), when comparing the FFE solutions from \cite{Komissarov2004b} 
with those ideal MHD solutions found in \cite{komissarov2005}. 
A similar observation was made in Ref.~\cite{Yang}, where the authors developed a family of analytic jet-like solutions and found the numerical evolution tends to a unique steady configuration (independently of initial data). 
They suggest (as observed previously in \cite{gruzinov2006,ruiz2012} in the context of neutron stars) that it might be the equatorial current sheet 
which determines the final state from the whole family of possible solutions.
%

The fact that the numerical resistivity is operating at the current sheet in our late time configurations
means the electromagnetic stress energy tensor is no longer conserved there;  
and thus, a source (or sink) of energy is expected to contribute when analyzing the total flux emitted (see expression \eqref{eq:stokes}).
As it was first pointed out in Ref.~\cite{Komissarov2004b}, the current sheet indeed supplies both energy and angular momentum to the force-free magnetosphere.

\begin{figure}
  \begin{center}
\includegraphics[scale=0.32]{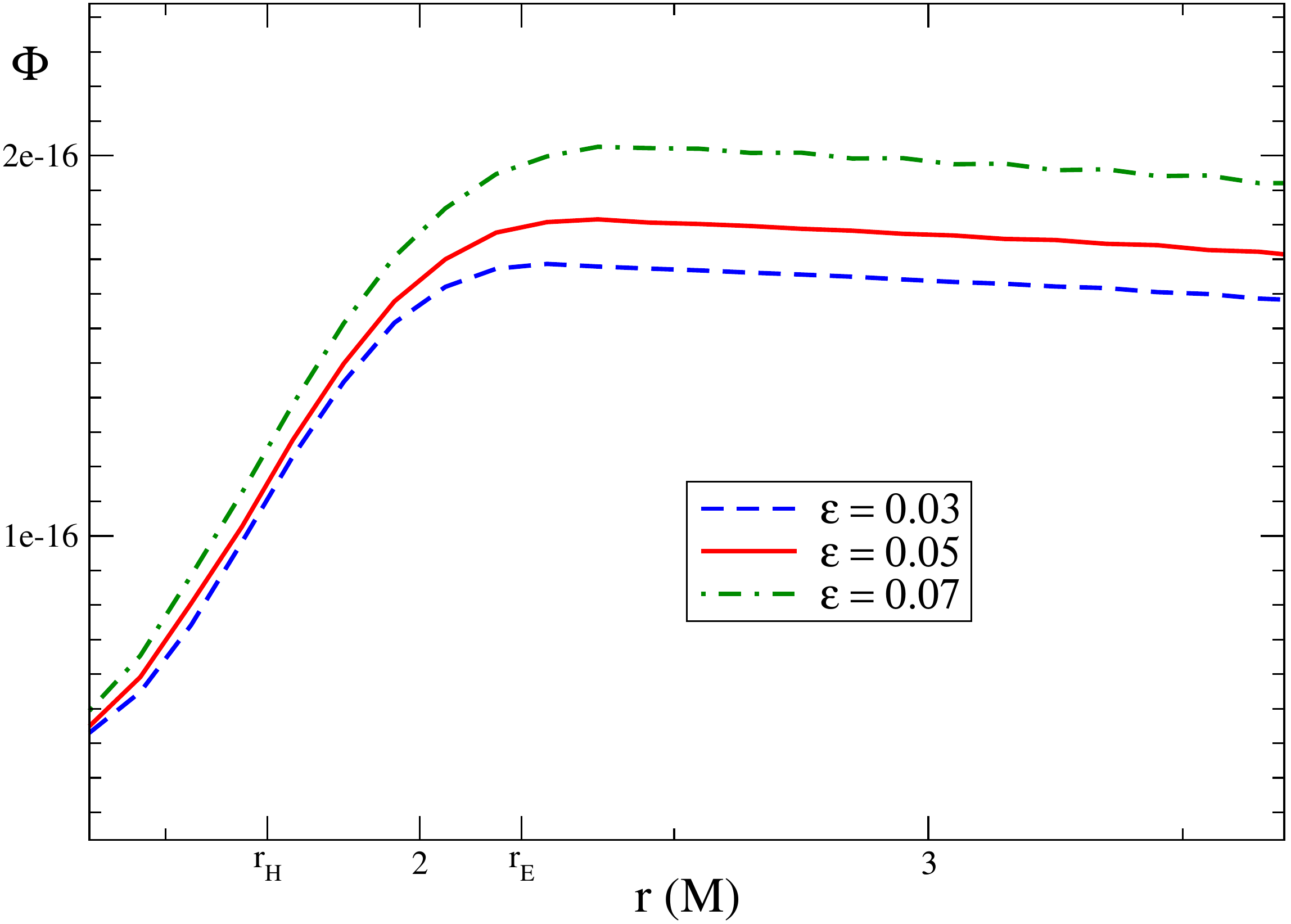}
  \caption{Net energy flux as a function of Cartesian radius: energy is being ``injected'' by dissipation at the current sheet.
  The flux is integrated on spherical layers at different radius, for representative numerical configurations (with $a=0.9$). The curves 
  corresponds to different values of the parameter $\varepsilon$, which controls the dissipative prescription (see Section \ref{sec:current_sheet}).\\
  $r_H$ and $r_E$ (horizon and ergosphere intersection with the equatorial plane) signals the region where energy is not conserved.
  }
 \label{fig:inyection} 
 \end{center}
\end{figure}

We have plotted, in Fig.~\ref{fig:inyection}, the net Poynting flux as it passes through $r=cst$ spheres, as a function of $r$. 
The figure illustrates how the flux increases considerably in between the horizon and the ergosphere.
The increment might be interpreted in part by negative energy falling into the black hole and in part as the energy of the electric field being dissipated at the current sheet; 
that energy is also negative, so its dissipation actually increases the energy as seen from infinity. 
Thus, we find that perplexing situation where dissipation acts as a positive source of energy.
The curves displayed corresponds to different values of the parameter controlling the mechanism to handle the current sheet, as described on Section \ref{sec:current_sheet}.
Essentially, the grater the value of the parameter $\varepsilon$ the more the electric field gets trimmed at the sheet and, hence, larger amounts of energy getting dissipated.
This reflects on the plot, which shows larger increments on the emitted power for larger values of $\varepsilon$.

\subsubsection{Dependence on boundary condition/location}

\begin{figure}[h!]
  \begin{center}
\includegraphics[scale=0.32]{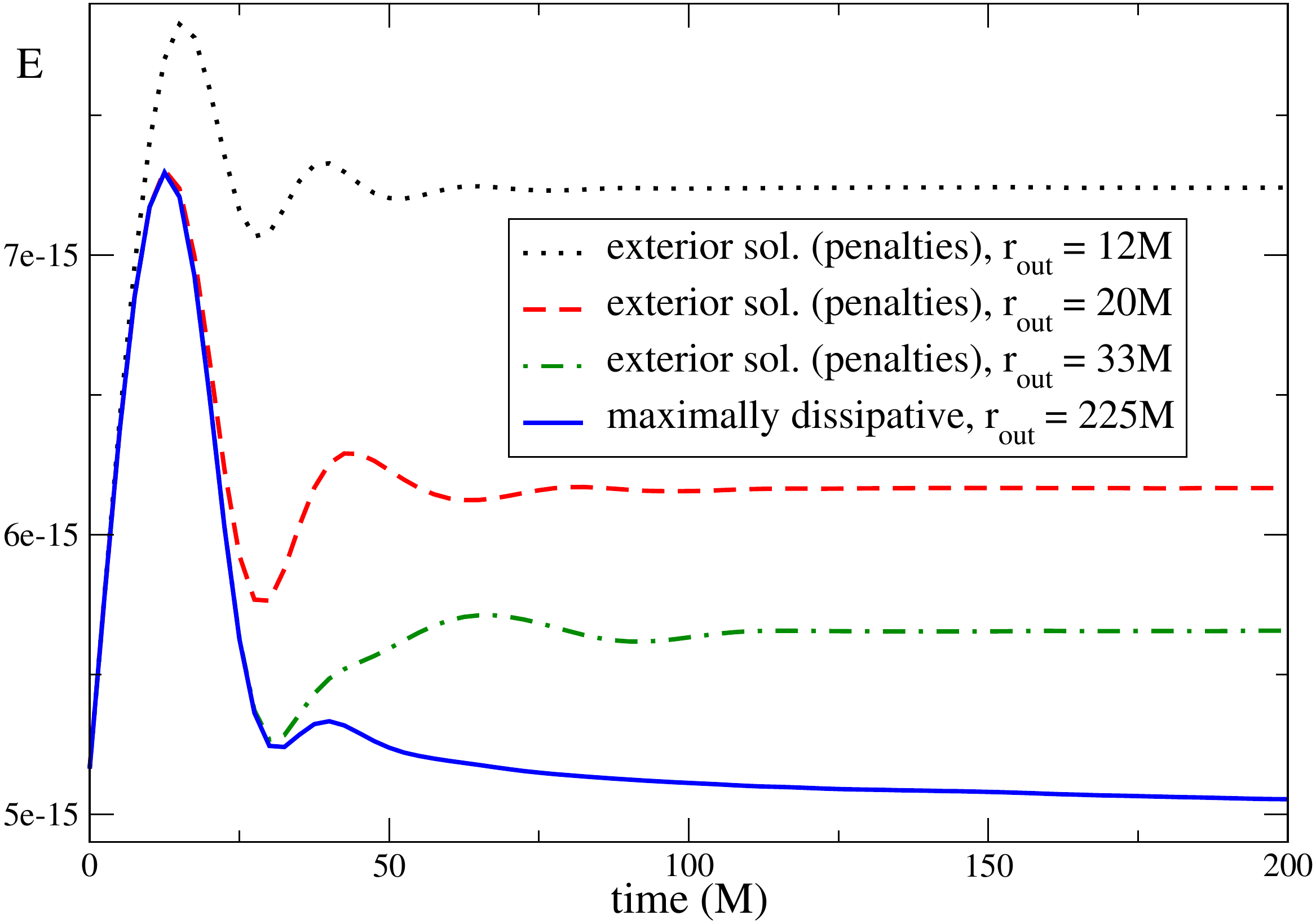}
\includegraphics[scale=0.32]{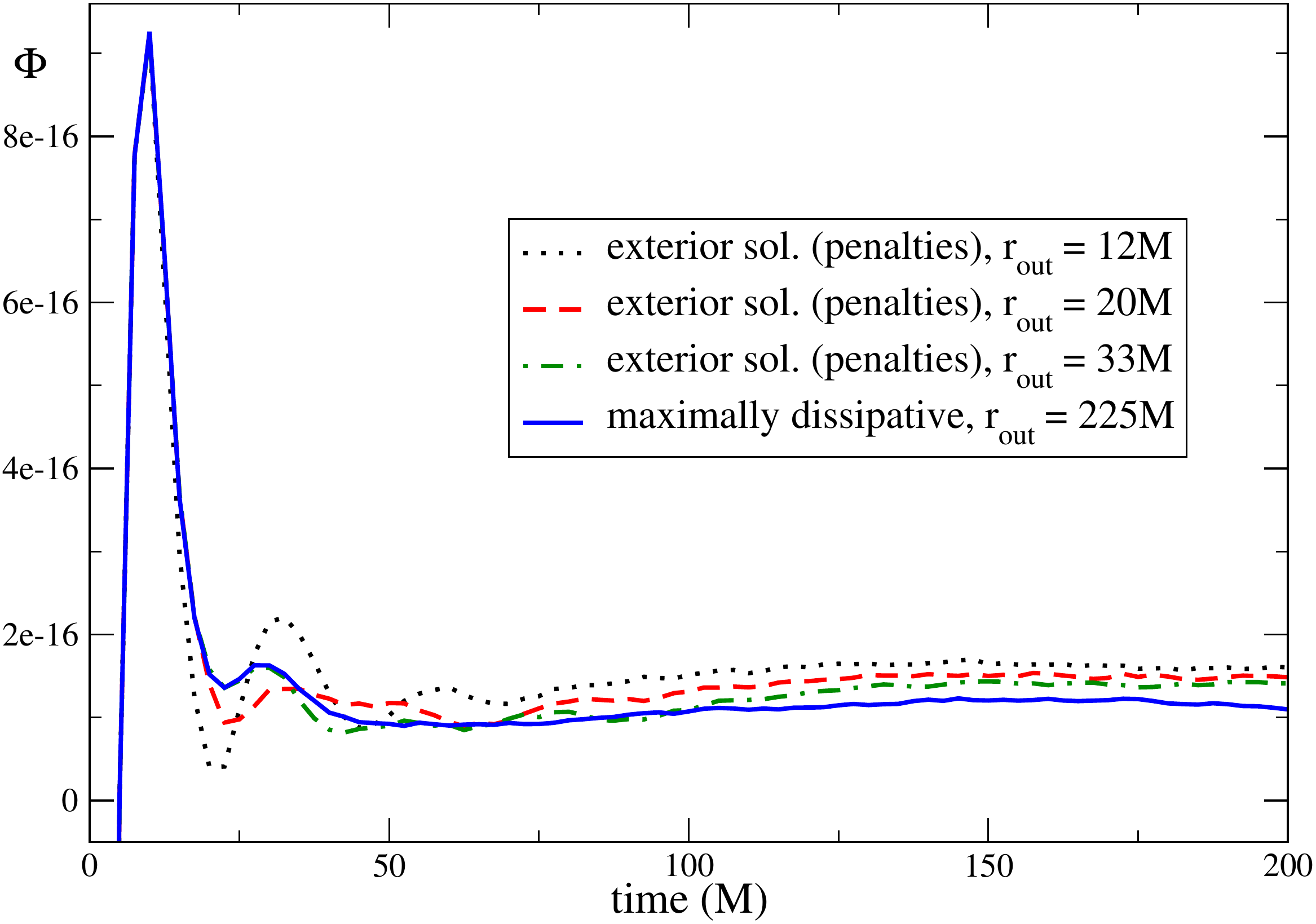}
  \caption{ Numerical solutions with an outer edge located at different radius, 
  further compared with a solution obtained under \textit{maximally dissipative} boundary conditions.
  \textbf{Top:} The evolution of the EM energy, when integrated up-to $r\sim 12M$. 
  \textbf{Bottom:} Net Poynting flux through the spherical layer at $r=3M$, as a function of time. }
 \label{fig:location}  
 \end{center}
\end{figure}

A natural question arises regarding our outer boundary condition treatment, which can be stated as follows: What is the influence the location of the numerical 
edge has on our late time solutions? To that matter, we have consider in Fig.~\ref{fig:location}, the evolution of the electromagnetic energy 
enclosed in a common region of space (specifically $r\in[1.35M, 11.65M]$), and the evolution of the net energy flux through a fixed spherical surface at $r=3M$, 
for simulations with their numerical edges placed at different radius. 
Another run using \textit{maximally dissipative} boundary conditions (and with the outer edge placed very far, $r_{out}\sim 225M$) 
was consider as well, for comparison.

Within our approach for the outer BC, we see the energy (top image) reaches an equilibrium on a time that depends on the location of the outer edge: 
when it is placed closer to the black hole, the solution equilibrates faster. The value attained also depends on the boundary location, being larger for smaller domains.
Whereas in the case of the maximally dissipative BC, the energy keeps slowly dropping towards the end of the simulation at $t=200M$. 
For the Poynting flux (bottom image), we notice 
that the net flux achieved under our BC, at different radial locations, seems to equilibrate and approach to a single value near the end 
(still slightly larger for smaller numerical domains). We further notice that it takes longer for the flux to equilibrate.
In the case of the outgoing boundary condition, we see the flux is slowly increasing after the first transient 
but then gradually starts to decrease at the final stage. 
Its value is, however, always lower when compared with the other numerical solutions. 
This observation might help on explaining the difference in the emitted power we found on Sec.~\ref{sec:spin_dep}. 

We have noticed that when imposing the outgoing boundary conditions through penalty terms, there is an initial perturbation
generated at the external surface, which propagates (at the speed of light) into the bulk.
When this perturbation reaches the central region, the solution gets ruined, and hence, the time we are allowed to run the simulation gets limited.
The reason for such perturbation to occur is that the penalties try to enforce the no-incoming condition from the very beginning, 
while the incoming modes associated to the initial data employed are nonzero.

\subsection{Misaligned Case}

\begin{figure*}[t]
  \begin{center}
  \begin{minipage}{8cm}
 \subfigure{\includegraphics[scale=0.21]{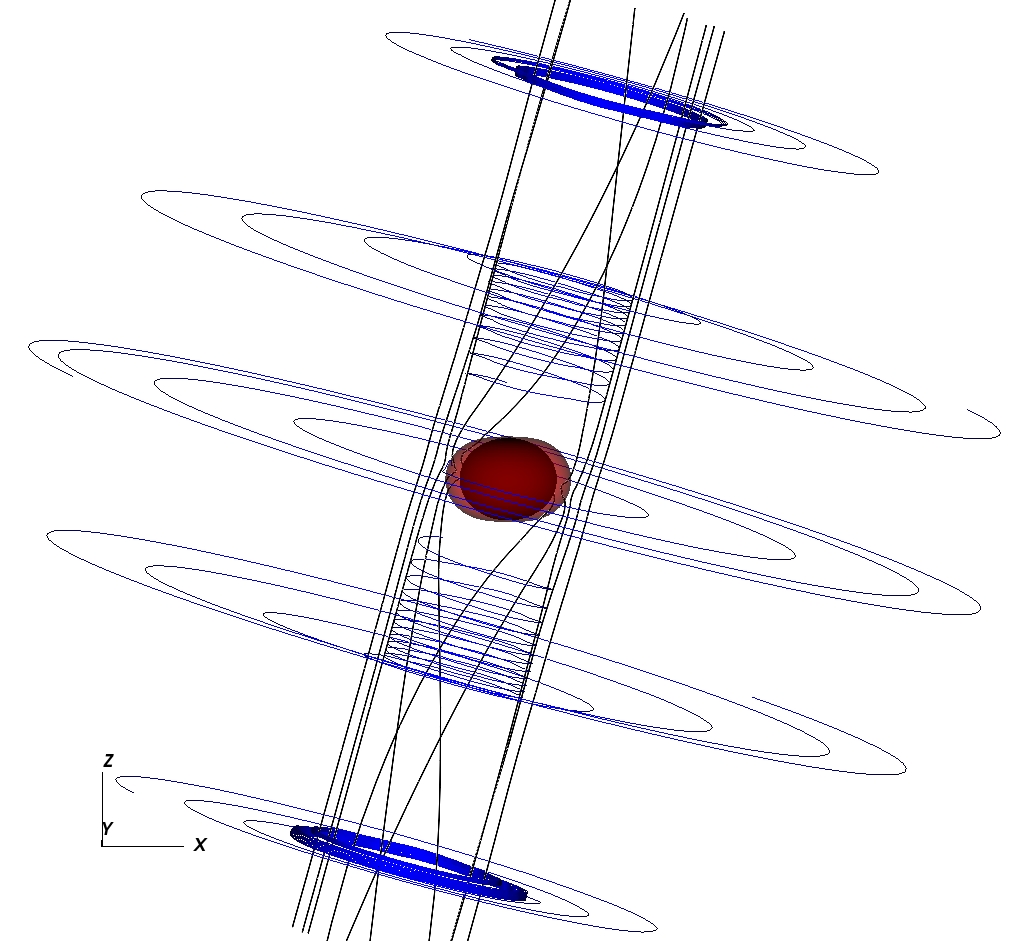}} 
\end{minipage}
\begin{minipage}{8cm}
\subfigure{\includegraphics[scale=0.26]{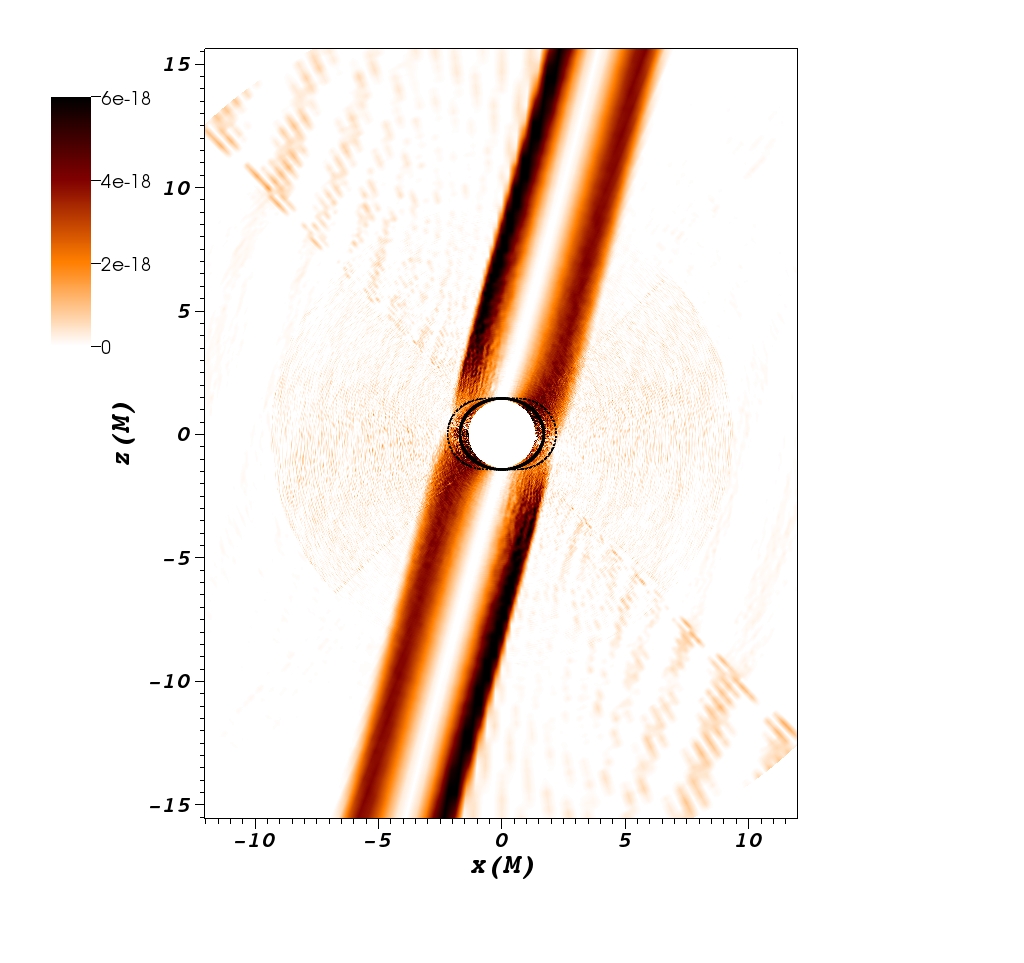}}
\end{minipage}
  \caption{Misaligned case ($\alpha_o = 15$\textdegree): late time numerical solution ($t=120M$) for a black hole with $a=0.9$. 
   \textbf{Left:} Representative streamlines of the magnetic field (black thick lines) and the electric field (blue lines).
   The solid black surface is the BH horizon, whereas the semitransparent (red) one represents the ergosphere.
   \textbf{Right:} The radial Poynting flux density is shown in color scale at the $x-z$ plane. 
   Thick solid and dotted black lines represents the black hole horizon and ergosphere, respectively.
   }
 \label{fig:PF_slice_misaligned} 
 \end{center}
\end{figure*}

We are interested on studying those cases in which the exterior magnetic field (generated by the accretion disk) 
is not aligned to the black hole rotational axis. 
We choose the $x-z$ plane for this displacement and denoted by $\alpha_o$ the angle between these two directions (see expr. \eqref{initial}).
A representative (high resolution) late time configuration, corresponding to a spin parameter $a=0.9$ and an inclination angle $\alpha_o = 15$\textdegree,
is displayed on Fig.~\ref{fig:PF_slice_misaligned}.
The electric and magnetic field attained (left panel image) are very similar to those in the aligned case, but now tilted. 
It is also apparent from the picture that the distribution of magnetic streamlines is no longer axi-symmetric.   
Again, a collimated Poynting flux is observed at the final stationary solutions (see right panel of Fig.~\ref{fig:PF_slice_misaligned}).
It can be seen that the jet follows the direction of the asymptotic (or exterior) magnetic field. 

\subsubsection{Dependence on inclination angle}

\begin{figure}[h!]
  \begin{center}
\includegraphics[scale=0.32]{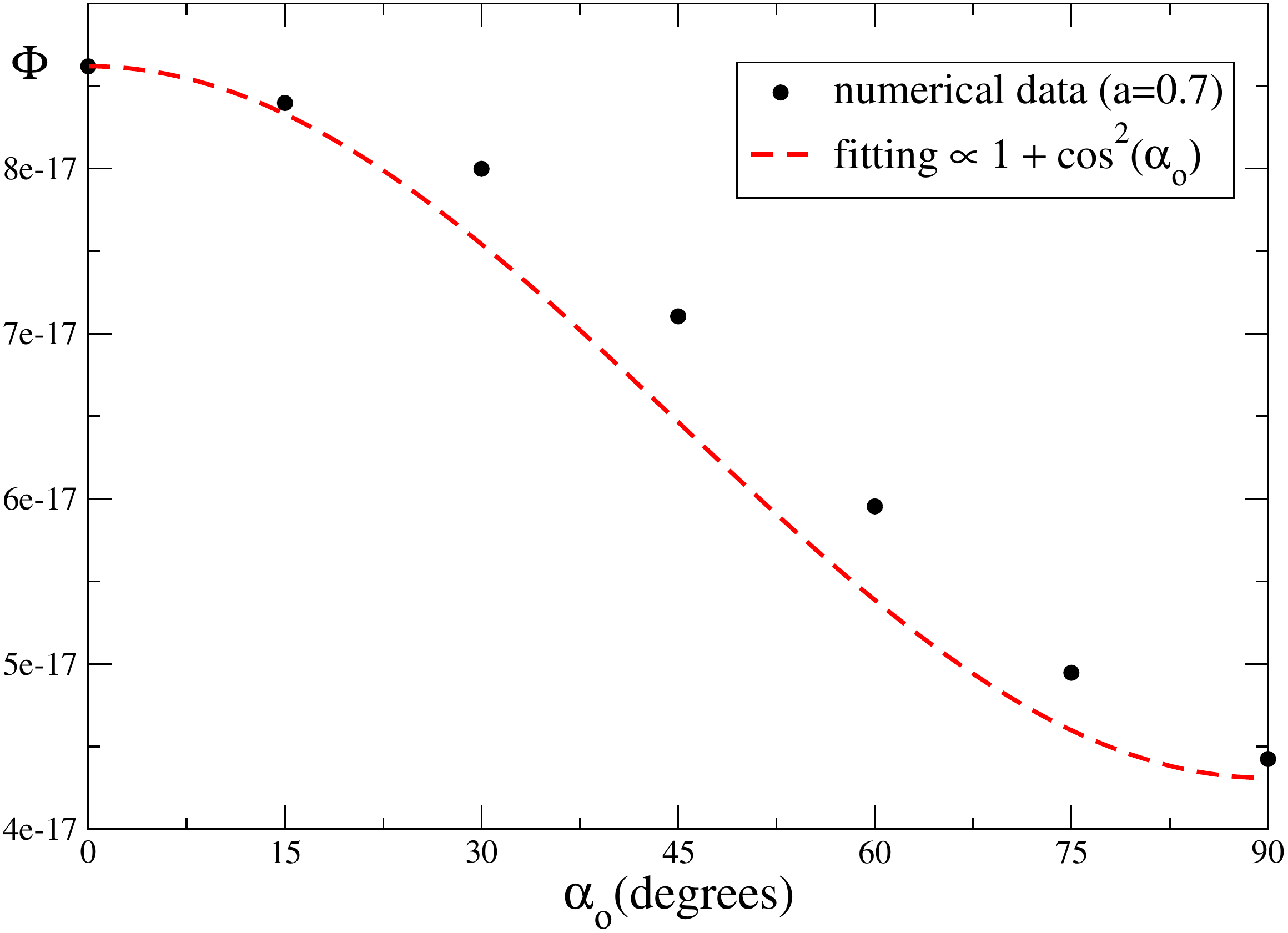}
  \caption{Dependence of the electromagnetic energy flux on the orientation of the asymptotic magnetic field.
   Net Poynting flux is plotted as a function of the inclination angle $\alpha_o$. 
   The dots correspond to the numerical values obtained, while the dashed (red) curve is a fit of the form $\Phi \propto 1+ \cos^{2}(\alpha_o)$.}
 \label{fig:PF_ang_dep} 
 \end{center}
\end{figure}

We now consider the dependence of the jet power on the inclination angle.
To that end, we vary the angle $\alpha_o$ from zero (aligned case) to $\frac{\pi}{2}$ (orthogonal case).
In Fig.~\ref{fig:PF_ang_dep}, this dependence is illustrated for a particular value of spin parameter ($a=0.7$)
and contrasted with the expected $\Phi \propto 1+ \cos^{2}(\alpha_o)$ behavior,
which is consistent with the one found for analytic non axi-symmetric jet solutions in \cite{gralla2015}. 
Again, our results compares well with those obtained in \cite{Palenzuela2010Mag} (figure 4)
and, as observed there, we see that even for the extreme situation where the two directions are orthogonal 
there is still a positive net electromagnetic energy flux. 
Moreover, the jet power decreases only to a fraction of the one in the aligned case.
In contrast, we do not find the small bump around $\alpha_o \approx 15$\textdegree reported in \cite{Palenzuela2010Mag}.
Instead, however, we find a similar departure of the trend starting from around $\alpha_o \approx 30$\textdegree.

As described in \cite{Palenzuela2010Mag}, we also find that the transversal structure of the magnetic field is given by 
a toroidal field with (counter)-clockwise rotation in the (anti)-aligned scenario, while in the orthogonal case ($\alpha_o  = \pi/2$), 
the system generates instead two counter-rotating toroidal fields off-set by a distance 
of about the black hole diameter (see figure 6 from Ref.~\cite{Palenzuela2010Mag}).


\section{Conclusions and Perspectives}

In this article, we have introduced a new finite-difference code to perform time-dependent 
and fully 3D numerical simulations of force-free electrodynamics on a Kerr background.
The code evolves, for the first time, the set of FFE equations developed in \cite{FFE}, which has improved properties in terms of well-posedness: 
it is a symmetric hyperbolic system and it remains so even when $\vec{E}\cdot \vec{B} \neq 0$,
while the more traditional FFE evolution equations has shown to be only weakly hyperbolic \cite{Pfeiffer,Pfeiffer2015, FFE}. 
The second and more important feature of the code consist on the implementation of stable and constraint preserving boundary conditions,
which allows to represent the numerical domain as embedded on a particular ambient configuration of the electromagnetic field.
Our treatment for the outer boundary relies on the penalty technique (already build-in for the multiple patch structure of the code),
that uses the characteristic information of the particular evolution system.
We chose a uniform magnetic field, as the surrounding environment, to set the incoming physical modes at the numerical boundary.
While for the constraints modes, we adapt an alternative method to avoid undesired constraint-violations to arise.
The inner numerical boundary, on the other hand, is always placed inside the black hole horizon and, therefore, 
no boundary condition has to be prescribed there.  

We have found stationary jet solutions that reproduces many of the central known results of the area. 
Namely: that all magnetic field lines crossing the ergosphere acquire a toroidal component (or angular velocity); 
that a collimated Poynting flux is generated, together with an electronic circuit at the surrounding plasma; 
and that energy is extracted from the black hole through the Blandford-Znajek mechanism.
We find a net energy flux dependence on black hole spin and on inclination angle 
(among the asymptotic magnetic field direction and the rotation axis) 
which is consistent with previous studies.

Our solutions are not just steady states, but truly stationary configurations, 
in the sense they achieve equilibrium --through boundary conditions-- with the ambient electromagnetic field chosen.
We have further analyzed the influence of the condition and location adopted for the outer boundary on our numerical results,
finding that the total emitted power might be larger when compared to analogue simulations that uses maximally dissipative boundary conditions.
The location of the outer edge, on the other hand, does not seem to significantly affect the late time configurations, including their luminosities. 

We remark that besides the Blandford-Znajek extraction mechanism acting in our late time solutions, 
there is energy injection at the equatorial current sheet \cite{Komissarov2004b}, where negative EM energy-at-infinity is being dissipated.
Moreover, we find the amount of energy incorporated to the stationary emitted flux is significant and depends on the method
to control the current sheet (see Sec. \ref{sec:current_sheet}). 
The more energy dissipated at the sheet, the larger the flux, which can reach almost twice the value observed at the black hole horizon.
Even though the process is a priori numerical (to avoid the break-down of the force-free approximation), 
it has been argued that when inertial effects are taken into account, the electromagnetic field is expected to transfer energy to the fluid
as to ensure a state of marginal screening of the electric field (i.e. $B^2 - E^2 \approx 0$), effectively dissipating EM energy.
And thus, in principle, the effect may be genuinely physical.
However, the details of this process are still unclear, and a finite-resistivity model 
to go beyond the force-free and ideal MHD regimes seems to be the essential step to determine a more precise jet structure and emitted power. 

In Appendix C, we have considered the monopole and split-monopole solutions as standard tests to our code, including a convergence analysis. 
We have also monitored the dynamical behavior of the constraints, ensuring no excessive deviations develops or enter the domain.

Our implementation of boundary conditions allows to place the outer numerical edge relatively close to the black hole horizon ($r_{out}\sim 10M$), 
where the force-free approximation is believed to be valid.
In this paper, we consider the surrounding environment to be a uniform magnetic field (with zero electric field), 
that corresponds to the magnetospheric Wald problem we wanted to study. 
However, we emphasize this ambient configuration might be easily replaced in the code to represent other physical scenarios;
even time-dependent ambient fields cases. 
One possibility, would be to refine the ambient field adopted, as to account for more particular (or realistic) accretion disk properties.
Nevertheless, we believe the vertical field contribution near the BH horizon should always appear as the dominant one, and thus, we would not expect the results to differ much from those obtained here.
A second --and more interesting-- possibility, is to consider the physical scenario of a black hole in translational motion through a magnetized plasma.
It has been first proposed in \cite{palenzuela2010dual,Luis2011}, that even a non-spinning black hole when moving relative to a plasma 
with an asymptotically stationary electromagnetic field topology, can produce jets. 
The problem has also been later approached analytically (see e.g. \cite{morozova2014,penna2015}). 
The idea is that, the kinetic energy is now what powers the jet, instead of the rotational energy of the black hole as in the usual BZ mechanism.
By appropriately boosting the uniform magnetic field configuration we use in this paper as initial/boundary data, it is possible to numerically implement the problem within our scheme.
The advantage of our numerical approach is that we can reach truly stationary solutions and, since we are in the frame of the black hole, we can --in principle-- probe
the whole range of possible boost velocities.
We are already exploring this scenario and hopefully we will present the results soon, in a forthcoming paper.

Finally, it would be also interesting to explore how the different physical modes behaves during the initial dynamical transient and how they set down towards the final configuration,
in both the magnetospheric Wald problem and on the boosted black hole scenario.
For doing it, we count with the projections (already built in the code) into the different characteristic subspaces of the system.


\section{Acknowledgments}

We would like to thank Luis Lehner for several very helpful discussions and orientations throughout the realization of this work.
We are also grateful to Miguel Meguevand and Gabriela Vila for many interesting comments and suggestions, specially at F.C. thesis defense. 

We acknowledge financial support from CONICET, SeCyT-UNC and MinCyT-Argentina.
This work used computational resources from CCAD Universidad Nacional de C´ordoba (http://ccad.unc.edu.ar/), in particular
the Mendieta Cluster, which is also part of SNCAD – MinCyT-Argentina.


\appendix

\section{Evolution Equations}

In a previous article, \cite{FFE}, we have constructed a well-posed evolution system for the theory of force-free electrodynamics.
These equations were written following a standard 3+1 decomposition of a background spacetime ($M$, $g_{ab}$), 
where the spacetime is foliated by the level surfaces of a smooth time function $t$, $\left\lbrace \Sigma_t \right\rbrace_{t\in \mathbb{R}} $ and  
an everywhere transversal vector field $t^a$ (see e.g. \cite{Thorne1982}). 
The normal vector to these surfaces is given by $n^a := -\alpha g^{ab} (dt)_b$, where the normalization factor $\alpha$ is known as the \textit{lapse function}.
And the departure of $t^a$ from the normal $n^a$ defines the \textit{shift vector}, $\beta^a := t^a - \alpha n^a$.
In adapted coordinates $\left\lbrace t, x^i \right\rbrace $ \footnote{For which the ``spatial'' coordinates $x^i$ are 
preserved along the vector field $t^a := (\partial_t)^a$ (\textit{time vector}).}
the line element reads,
\begin{equation}
ds^2 = (\beta^2 - \alpha^{2}) dt^2 + 2 \beta_i dx^i dt +h_{ij} dx^i dx^j
\end{equation}
where $h_{ij}$ is the intrinsic metric on the hypersurfaces $\Sigma_t$.

We adopt here a particular time-independent slicing of the Kerr spacetime (in which the stationary Killing vector field plays the role of $t^a$).
In particular, we consider the metric in the Kerr-Schild form and use the \textit{cubed sphere coordinates} for the angular directions.
In this way, $\partial_t \alpha = 0$, and the evolution equations found in \cite{FFE} can be slightly rewritten as:
\begin{widetext}
 \begin{eqnarray}
\partial_t \phi &=& \beta^k \partial_k \phi - \alpha \mathcal{D}_j B^{j}  - \alpha \kappa \phi - \frac{\alpha}{\Delta^2}\tilde{E}^k r_k  \label{dt_phi} \\ 
  \partial_t E^{i} &=& \left( \delta_{k}^{i} - \frac{\tilde{B}^{i}\tilde{B}_{k}}{\tilde{B}^2}\right) \left[ \beta^{k} \mathcal{D}_j E^{j} + \mathcal{D}_j (\alpha F^{kj}) \right]
 -  \frac{\alpha \tilde{S}^{i}}{\tilde{B}^2} \mathcal{D}_j E^j  
 + \frac{\tilde{B}^{i}}{\tilde{B}^2} \left[ \tilde{E}_{k} \mathcal{D}_j (\alpha F^{*kj}) - \tilde{E}_{\beta} \mathcal{D}_j B^{j} + r_{\beta} + \alpha \tilde{E}^k \partial_k \phi \right] \label{dt_E} \\ 
  \partial_t B^{i} &=&  \beta^i \mathcal{D}_j B^{j} - \mathcal{D}_j (\alpha F^{*ij}) +  \frac{\alpha}{\Delta^2}\hat{\epsilon}^{ijk} r_j \tilde{B}_k + \frac{\alpha \tilde{E}^i}{\Delta^2 \tilde{B}^2}\tilde{S}^k r_k - \alpha h^{ij}\partial_j \phi \label{dt_B}
\end{eqnarray}
\end{widetext}
where the electric and magnetic components of the electromagnetic tensor were defined from $n^a$ as,
\begin{eqnarray}
 E_a & := & F_{ab} n^b \\
 B_a & := & -F^{*}_{ab} n^b 
\end{eqnarray}
while the Poynting vector is given by,
\begin{equation}
S^a := n_e \epsilon^{eabc}E_b B_c
\end{equation}
And the two electromagnetic invariants are denoted,
\begin{equation}
 G:= F^{ab}F^{*}_{ab} \quad \text{ ; } \quad F:= F^{ab}F_{ab}
\end{equation}

We recall here some other important definitions:
\begin{eqnarray}
 \tilde{E}^i &=& E^i + \sigma B^i \nonumber\\
 \tilde{B}^i &=& B^i - \sigma E^i \nonumber\\
 \tilde{S}^i &=& (1+\sigma^2 ) S^i \nonumber
\end{eqnarray}
\begin{equation}
 \sigma := \frac{\sqrt{F^2 + G^2}-F}{G} 
\end{equation}
\begin{equation}
 \Delta^2 := \tilde{B}^2 - \tilde{E}^2 
\end{equation}
\begin{equation}
  r_i  := \frac{\alpha^2}{4} \left( \partial_i (G/\alpha^2 ) + \sigma \partial_i (F/\alpha^2 )  \right) 
\end{equation}
Also, we have denoted $\hat{\epsilon}^{ijk} \equiv n_a \epsilon^{abcd}$
(the induced volume element on the hypersurface), and $\mathcal{D}_j (\cdot) \equiv \frac{1}{\sqrt{h}} \partial_j (\sqrt{h} \text{  } \cdot \text{  })$\\
Naturally, $F^{ij}$ and $F^{*ij}$ can be written in terms of electric and magnetic fields through the relations,
\begin{eqnarray}
\alpha F^{ij}  =  E^{i} \beta^{j} - E^{j} \beta^{i}  + \alpha \hat{\epsilon}^{ijk} B_{k} \nonumber\\
\alpha F^{*ij} = \beta^{i} B^{j}  - B^{i}\beta^{j}  + \alpha \hat{\epsilon}^{ijk} E_{k} \nonumber
\end{eqnarray}

Finally, we incorporate to the evolution system a damping strategy taken from \cite{alic2012} to have a better control on the algebraic constraint $G=0$. That is,
\begin{equation}
 \partial_t E^i \rightarrow \partial_t E^i  - \alpha ~ \delta ~ \frac{(E_k B^k)}{B^2} B^i \label{damp}
\end{equation}
Since our equations mathematically preserve this constraint during the evolution\footnote{As compared to those evolution systems which does not include the complete force-free current 
(like e.g. \cite{Komissarov2004b, spitkovsky2006, Palenzuela2010Mag, parfrey2012, alic2012}).}, we can adopt a moderate coefficient $\delta \sim 100$
to enjoy the constraint cleaning proprieties, while at the same time --as pointed out in \cite{Pfeiffer2015}--, avoid the complications of having stiff terms
which would demand the use of implicit-explicit schemes (like those used on \cite{alic2012}). 
Also notice that the damping term does not alter the characteristic structure of the original evolution equations \eqref{dt_phi}-\eqref{dt_E}-\eqref{dt_B}.

\section{Conservations, fluxes, and the Blandford-Znajek mechanism}

\begin{figure}
  \begin{center}
\includegraphics[scale=0.45]{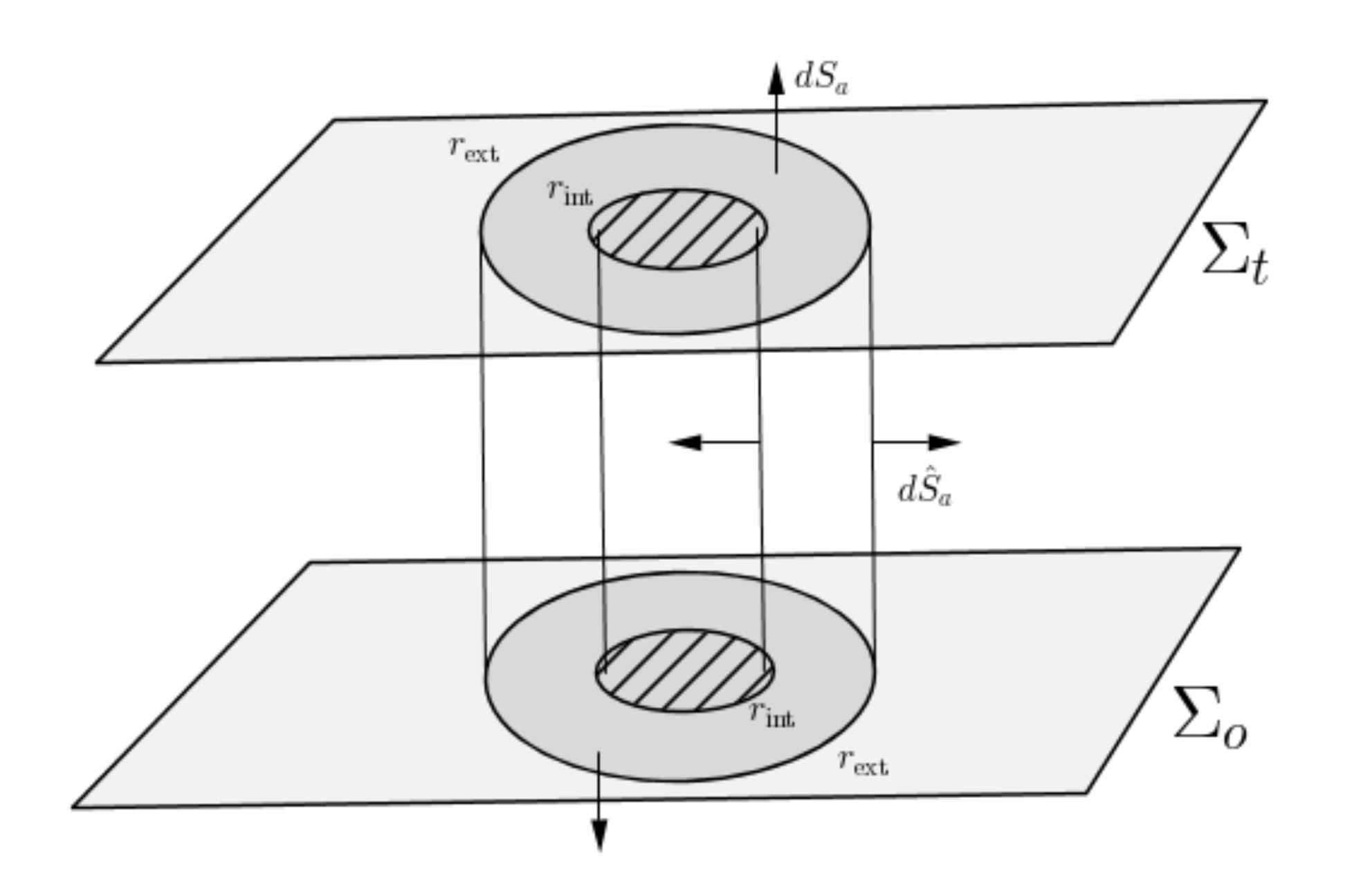}
  \caption{Stokes theorem: the total flux of a conserved current, through the boundary of a spacetime region $S\subset M$, is zero. }
\label{fig:conserv}
 \end{center}
\end{figure}

From a conserved stress tensor $T^{ab}$ and a Killing vector field $\xi^{a}$, 
we can build a conserved current $j^{a}:=T^{ab} \xi_b $. Applying Stokes theorem,
\begin{eqnarray}
 0 &=& \int_{S} (\nabla_a j^a ) \sqrt{-g} \, d^4 x = \int_{\partial S} j^a dS_a \label{eq:stokes}\\
   &=& \int_{\Sigma_t} j^a dS_a - \int_{\Sigma_o} j^a dS_a + \int_{r_{ext}} j^a d \hat{S}_a - \int_{r_{int}} j^a d \hat{S}_a \nonumber
\end{eqnarray}
One may interpret this as the existence of a quantity (defined on the hypersurfaces  $\Sigma_t$) which is preserved
up to fluxes that enters or leaves the region $S$ (see Fig.~\ref{fig:conserv}).

If the region $S$ is enclosed by two time slices and two spherical ($r=cst$) surfaces, the boundary normals are:  
\begin{eqnarray}
 && dS_a = - (dt)_a \sqrt{-g} \, d^3 x \nonumber\\  
 && d\hat{S}_a = (dr)_a \sqrt{-g} \, dt \, d^2 x \nonumber 
\end{eqnarray}

The force-free condition implies that the electromagnetic stress tensor is conserved. 
The Kerr spacetime has two Killing vector fields, $k^a$ and $m^a$, related with stationarity and axial symmetries, respectively.
The associated conserved currents are identified with the electromagnetic energy and angular momentum.
We define first the four-momentum,
\begin{equation}
 p^a := - T^{ab} k_b
\end{equation}
The electromagnetic energy at time $t$ is given by,
\begin{equation}
 E(t):= \int_{\Sigma_t} \mathcal{E} \sqrt{h}\, d^{3}x 
\end{equation}
with $\mathcal{E}:= -p^a n_a $ being the energy density. 

According to Stokes theorem, we see that the energy difference on a time interval would be given by the net flux of energy across the
inner and outer spherical surfaces (during this interval). That is,
\begin{equation}
 \Delta E \equiv E(t) - E(t_o ) = \Phi(r_{ext}) - \Phi(r_{int}) \nonumber 
\end{equation}
\begin{equation}
 \Phi(R) := \int_{R} \sqrt{-g} ~ p^r  \, dt \, d^2 x \label{flux-density} 
\end{equation}
where $ p^r $ is the radial Poynting flux density, and $\Phi(R)$ represents the flux through a sphere of radius $R$ during the time interval.

In a stationary situation, the energy would remain constant and, hence, the flux of electromagnetic energy through a spherical shell will be the same at any radius.
For an axially symmetric field configuration, on the other hand, the flux of energy through the black hole horizon can be computed (see equation (30) in \cite{Palenzuela2010Mag}) and gives, 
\begin{equation}
 \phi_{\mathcal{E}} |_{r=r_{H}} = 2 (B^{r})^2 \, r_{H} \, \Omega_F \, (\Omega_H - \Omega_F ) \sin^2 \theta \label{BZ-cond}
\end{equation}
Therefore, if $B^r \neq 0$ and $0<\Omega_F < \Omega_H$ (to which we will refer as the BZ conditions \cite{Blandford}), there would be an outward directed flow of energy from the horizon to the asymptotic region.
This is interpreted as rotational energy from the black hole being extracted electromagnetically. 
While the Maxwell field is actually falling into the black hole, its associated Killing energy is negative (which is possible inside the ergoregion), and thus, 
seen as a positive flux of EM energy by a ``distant observer''.\\

Analogously, but for the axial Killing field $m^{a}$, one can define the angular momentum at time $t$, 
\begin{equation}
 L(t):= \int_{\Sigma_t} S_a m^a \sqrt{h} \, d^{3}x 
\end{equation}
with, $ T^{r}_{\phantom a a} m^a $, defining the radial angular momentum flux density. 

\section{Numerical Tests}

In this appendix we present further tests to our numerical implementation.
We first consider the monopole and split-monopole approximate solutions for slowly rotating black holes,
configurations commonly used to test FFE codes (see e.g. \cite{Komissarov2004b, mckinney2006, Palenzuela2010Mag, shapiro2013}).
Then, we analyze convergence by considering the zeroth-order approximation to the monopole solution.
Finally, we monitor the dynamical behavior of the constraints to see they are properly preserved during the evolution.

\subsection{Known force-free solutions}

\subsubsection{Monopole solution}

Blandford and Znajek \cite{Blandford} has constructed a solution, in the limit of a slowly rotating black hole, that represents a magnetic monopole
matching the flat spacetime solution of Michel \cite{michel1973} at large radius and satisfying Znajek's condition \cite{znajek} at the event horizon.
This analytic configuration predicts $\Omega_F = 0.5 ~ \Omega_H$, and leads to a toroidal magnetic component
\begin{equation}
 B_{\phi} = -\frac{1}{8} a B_0 \sin^2 \theta + O(a^{3})
\end{equation}

\begin{figure}
  \begin{center}
\includegraphics[scale=0.32]{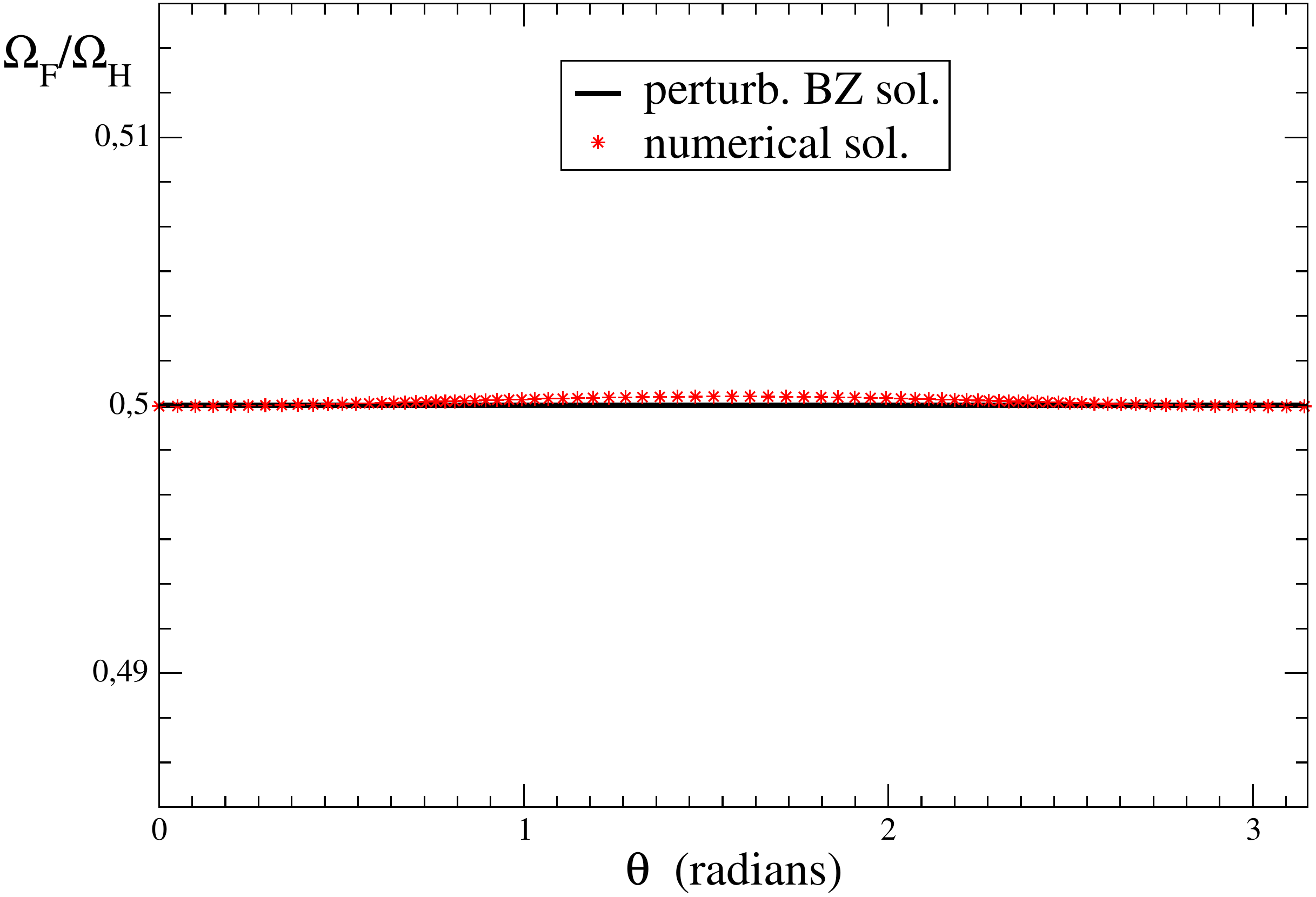}
\includegraphics[scale=0.32]{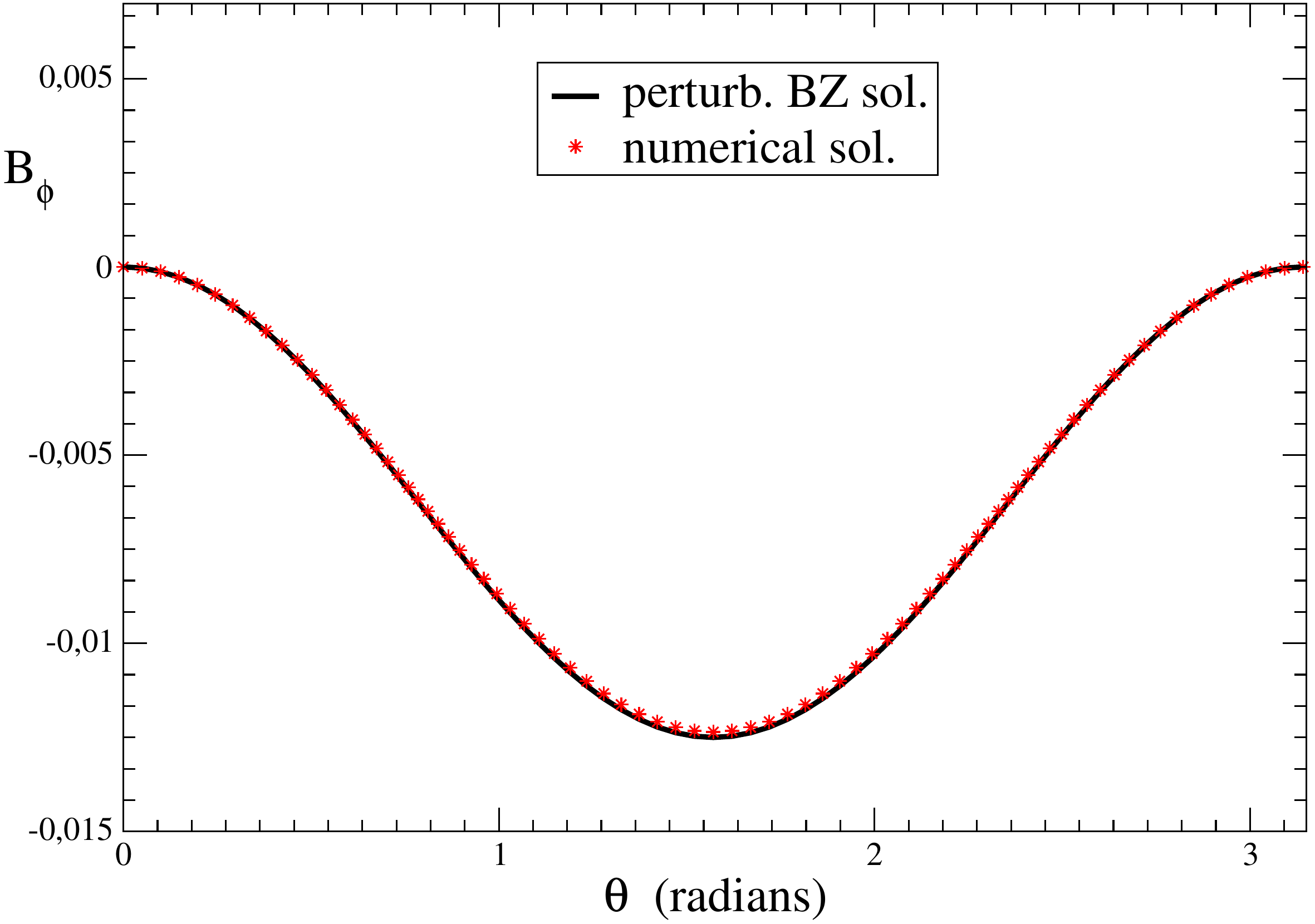}
  \caption{ Monopole magnetic field of strength $B_o = 1$, for a black hole spin $a=0.1$ at $t = 100M$. 
  \textbf{Top:} Angular frequency of magnetic field lines at $r = 3M$. 
  Comparison of our numerical results with the perturbative BZ solution which has $\Omega_F = 0.5 \Omega_H$.
  \textbf{Bottom:} Angular distribution of the toroidal component of the magnetic field (at $r = 3M$).
  The numerical solution matches very well the perturbative one. }
 \label{fig:monopole_ang_dep} 
 \end{center}
\end{figure}

In spite of being unphysical, is one of the simplest configurations showing the extraction of energy 
through the BZ mechanism and the first force-free solution tested numerically \cite{komissarov2001}.
We employ here, as initial data, the purely radial zero-order solution,
\begin{equation}
 B^r = \frac{B_0}{\sqrt{h}} \sin \theta
\end{equation}
adopting $B_0 = 1$ (in geometrized units) and $a=0.1$. 
We use six grids to cover a region between $r=1.9M$ and $r=9.9M$ (as seen in Sec. \ref{sec:grids}), 
with 160 points in each of the angular and radial directions.
The numerical solution evolves towards the perturbative Blandford-Znajek solution.
Fig.~\ref{fig:monopole_ang_dep} reflects this fact by showing the angular distributions of 
$\Omega_F / \Omega_H$ and $B_{\phi}$, at $r=3M$.
This figure is comparable to fig.~2 of \cite{Komissarov2004b} and fig.~3 of \cite{mckinney2006}.

We notice that, as mentioned in references \cite{Komissarov2004b,mckinney2006}, the final state of the numerical evolution
is insensitive to the details of the initial condition. 
Moreover, our results seems to be also insensible to the choice of outer boundary condition
among those that correspond to the zero and first order\footnote{See equations \eqref{split:1}--\eqref{split:3}, below.} approximate solutions. 

\subsubsection{Split-monopole solution}

\begin{figure}
  \begin{center}
\begin{minipage}{4.2cm}
 \subfigure{\includegraphics[scale=0.145]{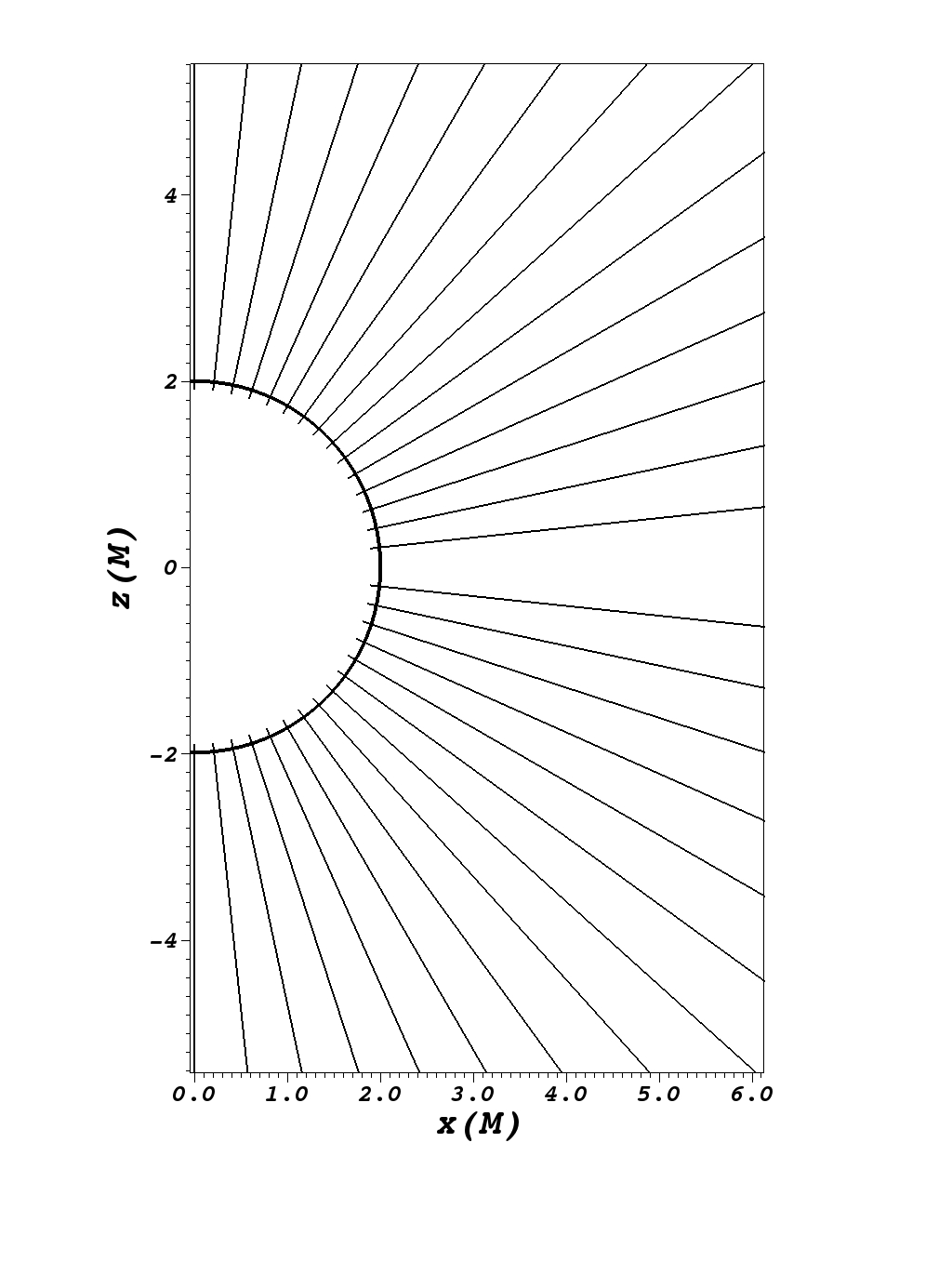}} 
\end{minipage}
\begin{minipage}{4.2cm}
\subfigure{\includegraphics[scale=0.145]{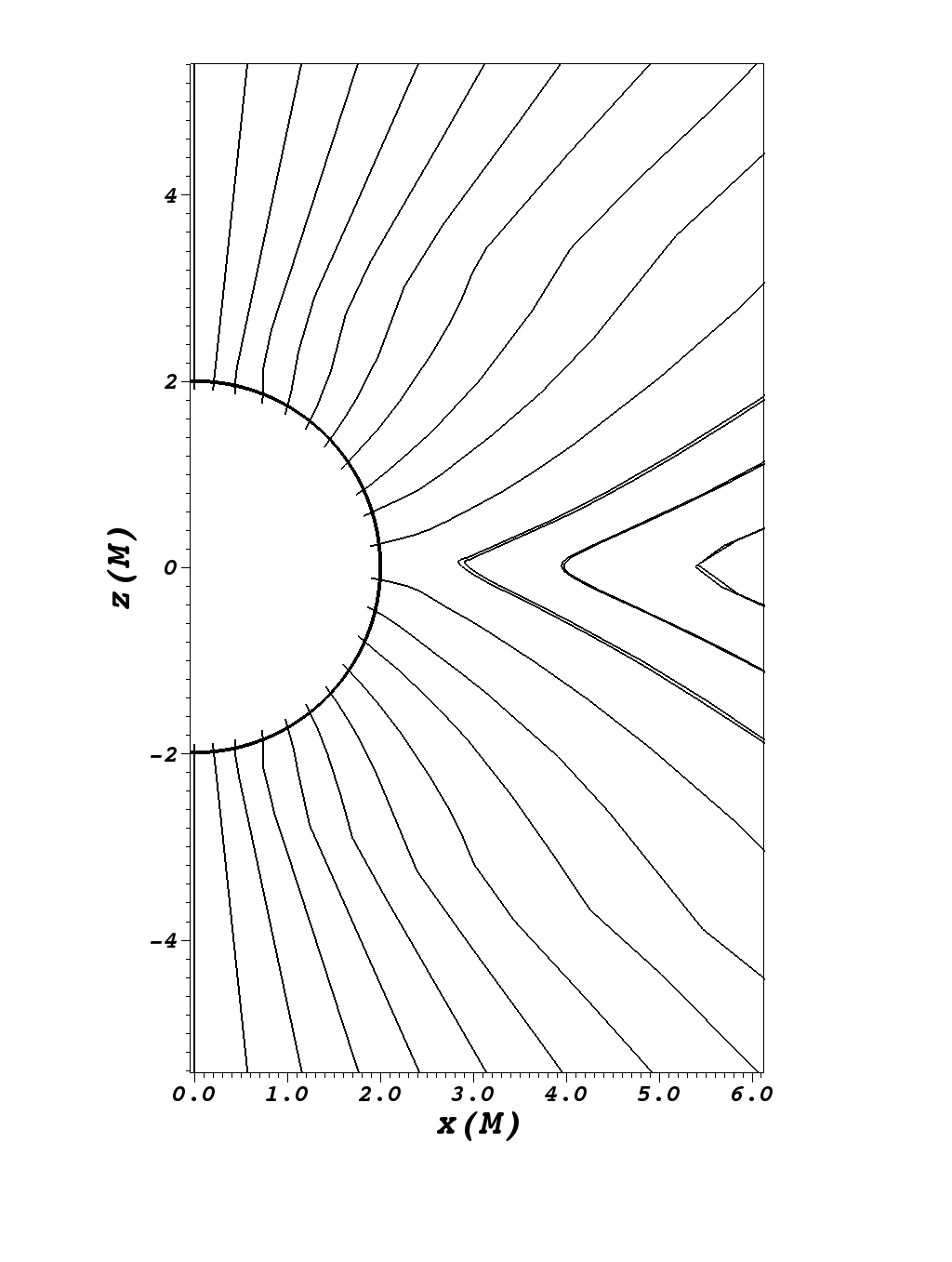}}
\end{minipage}
  \caption{ Split monopole field of strength $B_o = 1$ for a black hole spin $a=0.1$. 
  Poloidal magnetic field lines at the $x-z$ plane around the black hole horizon (thick half circle). 
  \textbf{Left:} Initial split-monopole configuration at $t=0$.
  \textbf{Right:} Configuration of the numerical solution at $t=5M$. The magnetic field lines reconnect and are spelled from the central region. }
 \label{fig:split} 
 \end{center}
\end{figure}

Alternating the sign of the magnetic field on one of the hemispheres leads to the ``split-monopole'' solution.
It is still a force-free solution but with zero total magnetic flux, and hence, more relevant from a physical perspective. 
As pointed out in \cite{Komissarov2004b}, this configuration is sensible only if there is a perfectly conducting disk in the equatorial plane of the black hole.
Otherwise, the equatorial current sheet cannot be stable: the magnetic fields lines will reconnect and be pushed away.
This is what happens in our numerical simulations, as illustrated in Fig.~\ref{fig:split}.

We take, as initial/boundary data, the first order accurate split-monopole solution in Kerr-Schild coordinates from \cite{shapiro2013} (see also \cite{mckinney2004}),
\begin{eqnarray}
 && B^r \simeq B_0 \frac{\alpha M^2}{3r^2}\left( 1- \frac{a^2}{r^2}\cos \theta \right) \label{split:1}\\
 && B^{\phi} \simeq -B_0 \frac{a \alpha M}{24r^2} \left( 1 + \frac{4M}{r}\right) \label{split:2} \\
 && E_{\theta} \simeq -B_0 \frac{a}{24\alpha} \left[ 1 +  \frac{2M \alpha^2}{r}\left( 1+\frac{4M}{r} \right)\right] \sin \theta \label{split:3}
\end{eqnarray}
with $B^{\theta} \simeq 0 \text{ ,  } E_r \simeq 0 \text{ ,  } E_{\phi} \simeq 0 $.
We adopt $B_0 = 1$ for one hemisphere and $B_0 = -1$ for the other. And again, we consider the slowly rotating black hole with $a=0.1$.

Fig.~\ref{fig:split} is in good agreement with fig.~3 of \cite{Komissarov2004b} and fig.~2 of \cite{shapiro2013},
showing rapid reconnection of the magnetic field lines. 
In Refs. \cite{mckinney2006, shapiro2013}, the authors has considered a different prescription for resistivity at the current sheet,
by ``nulling the inflow velocity'' at the equatorial plane.
This way, they were able to avoid magnetic reconnection and retain the split-monopole configuration for longer periods of time.


\subsection{Convergence}

We test convergence using the initial/boundary data of the monopole solution on a slowly rotating ($a=0.1$) Kerr black hole.
We consider a numerical domain between $r=1.9M$ and $r=5.9M$ (made from six grids) with three different spatial resolutions, at the ratio $4:2:1$.
The coarsest resolution uses $21\times21\times41$ points for each of the six grids, while the finest one has $81\times81\times161$.

We compute the \textit{precision factor} $p$ as,
\begin{equation}
 p \sim \log_2 \left( \frac{||u^{(1)} - u^{(2)}||_2 }{||u^{(2)} - u^{(4)}||_2 } \right) 
\end{equation}
with ($u^{(1)}$, $u^{(2)}$, $u^{(4)}$) being the numerical solutions obtained with spatial steps ($\Delta x$, $\Delta x/2$, $\Delta x/4$), respectively.

\begin{figure}
  \begin{center}
\includegraphics[scale=0.32]{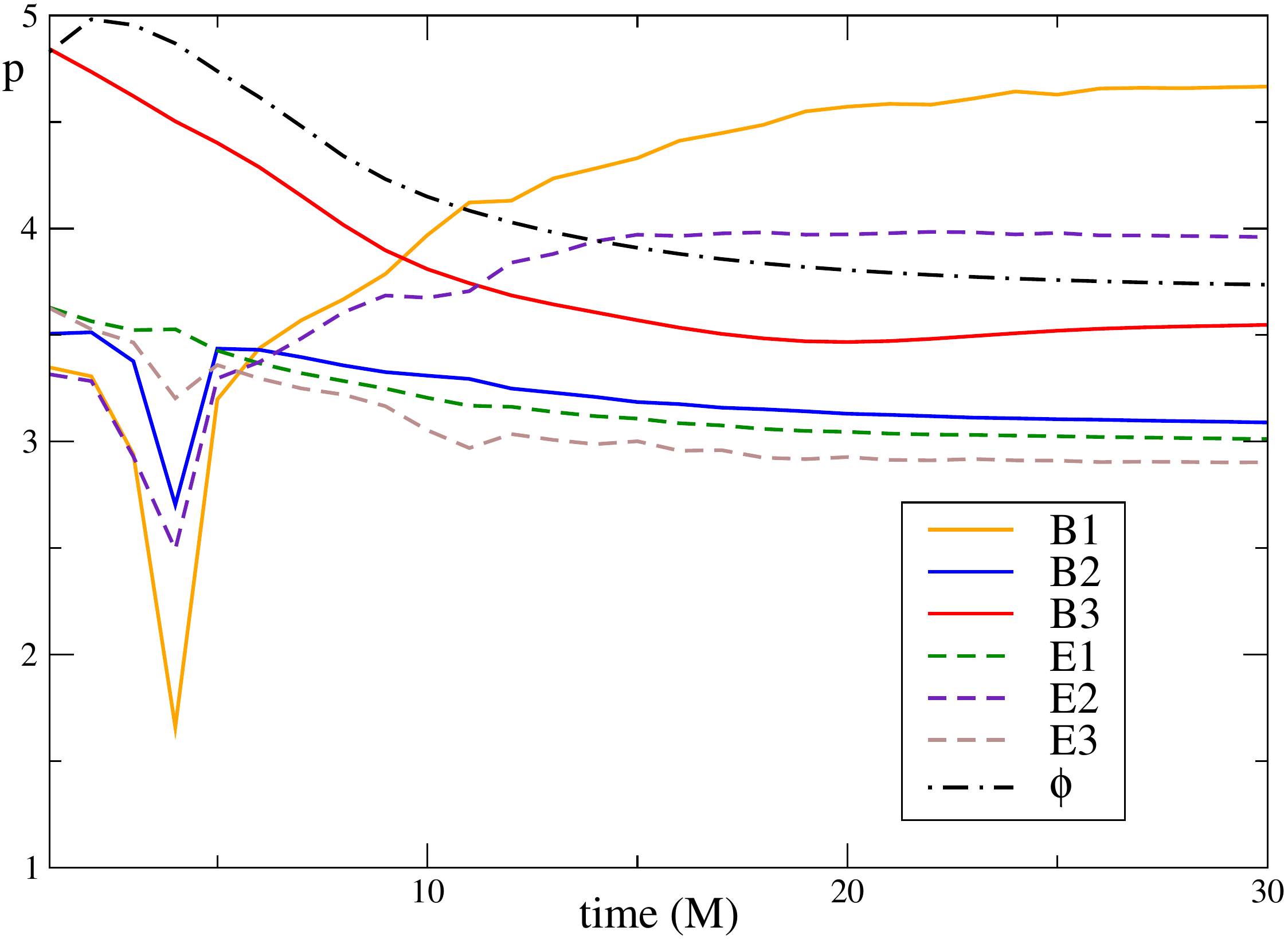}
\includegraphics[scale=0.32]{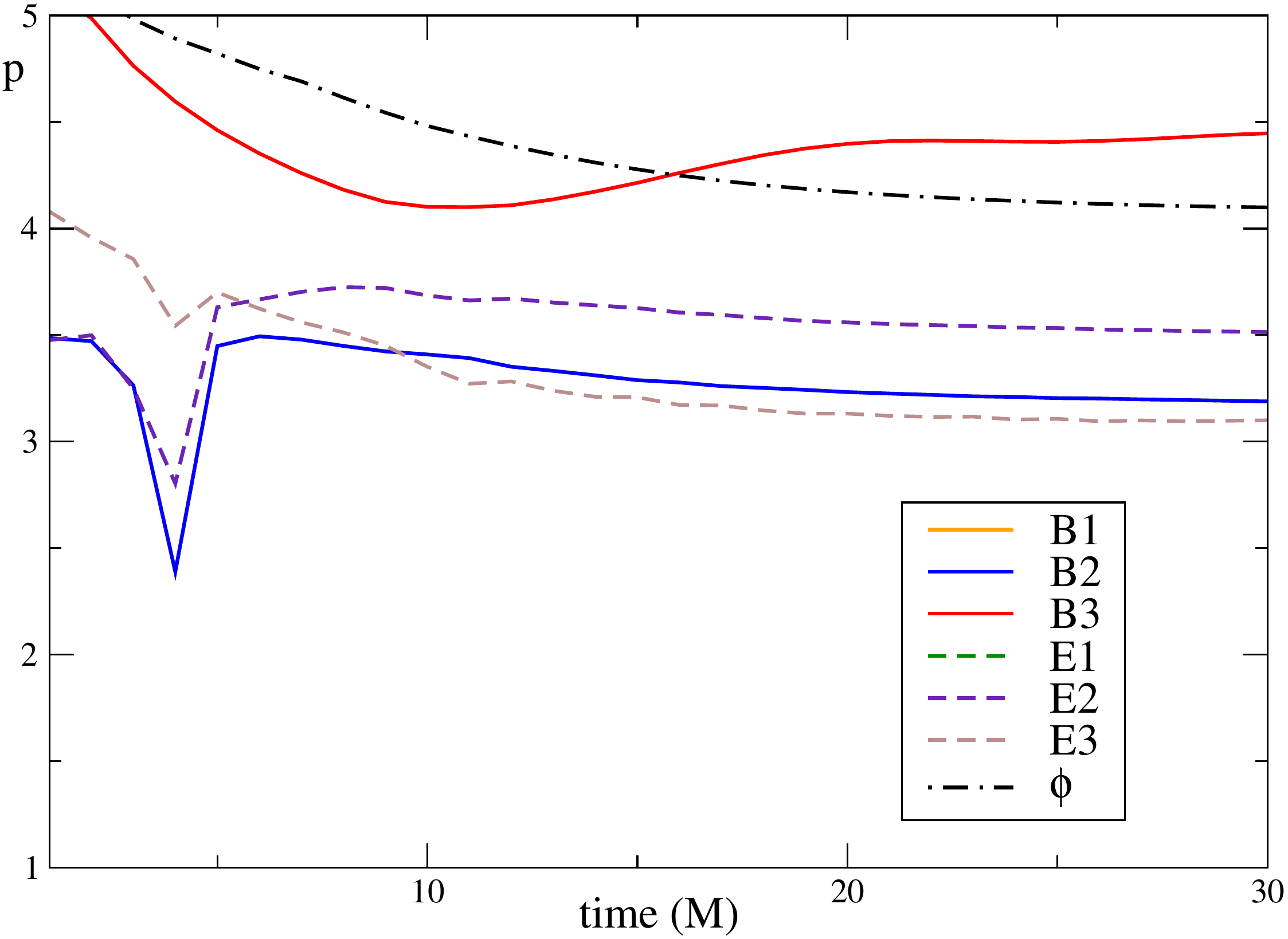}
  \caption{ Evolution of the precision coefficient, $p$, for the monopole solution (with $a=0.1$), for each of our dynamical variables. 
  It is computed for two representative grids: one over the equator (\textbf{top}) and one at the polar region (\textbf{bottom}).}
 \label{fig:conv_mono} 
 \end{center}
\end{figure}

In Fig.~\ref{fig:conv_mono}, we plot the precision factor as a function of time for all our dynamical variables.
As can be seen, most of them converge to $p\sim 3$, which agrees with the precision of the difference operator at the boundaries (the lowest order of the whole scheme).
The rest of the fields are around a value $p\sim 4$, corresponding to the precision of the Runge-Kutta method we use for time integration.
Notice in the bottom image of Fig.~\ref{fig:conv_mono}, the curves of the components B1-B2 and E1-E2 are superimposed, which can be easily understood by symmetry considerations.

\begin{figure}[h!]
  \begin{center}
\includegraphics[scale=0.32]{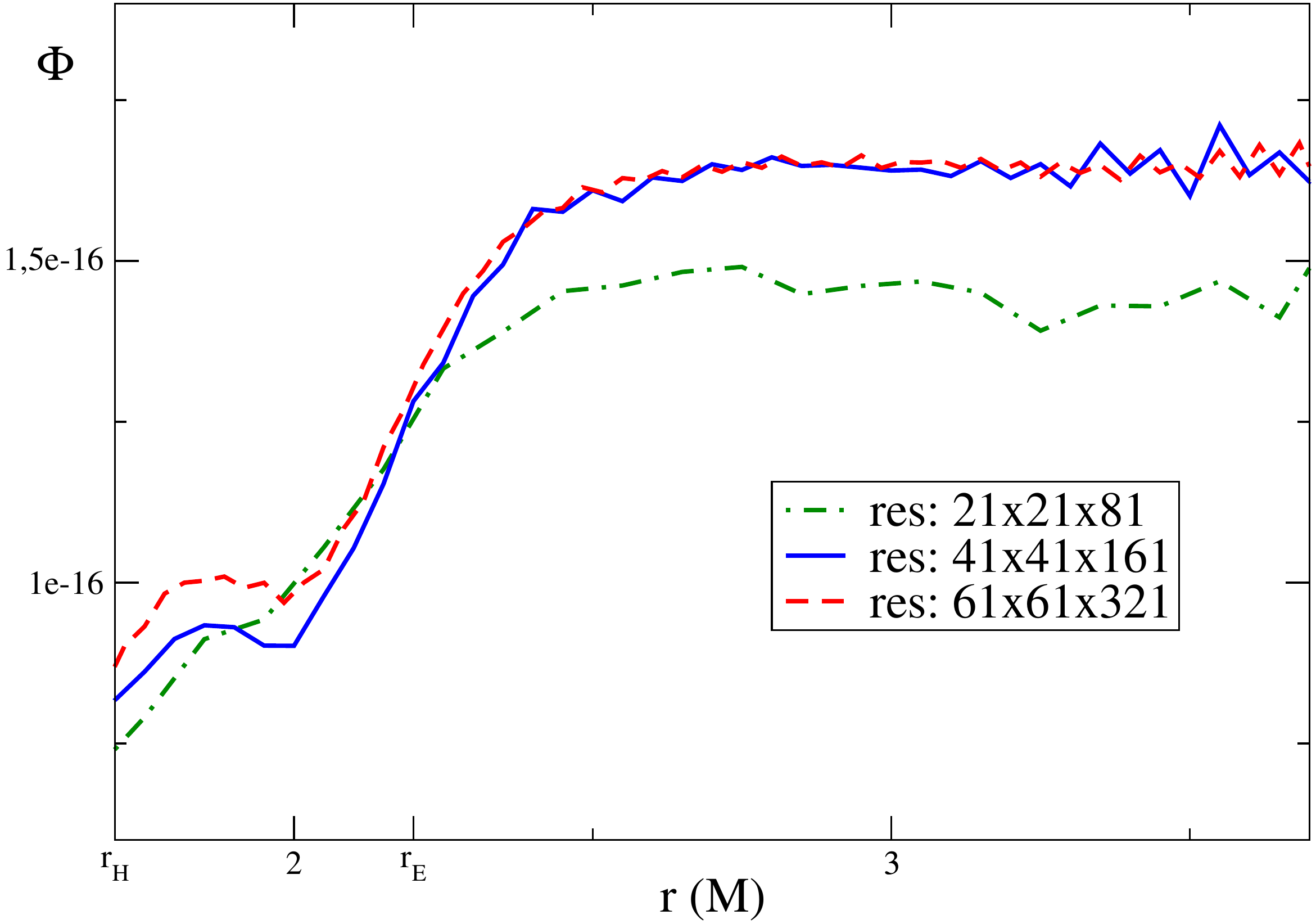}
  \caption{Net energy flux as a function of Cartesian radius for the late time solutions.
  The curves corresponds to three different resolutions, for BH spin $a=0.9$ and outer boundary located at $r_{out}=17.35M$.
  $r_H$ and $r_E$ represents the horizon and ergosphere intersection with the equatorial plane. }
\label{fig:conv_gral}
 \end{center}
\end{figure}

For the cases considered in Sec.~\ref{sec:results}, in which there is dissipation taking place at the current sheet,
we wont expect the code to converge. Indeed, the magnetic field becomes discontinuous there and our finite difference schemes can not deal appropriately with it.  
However, in Fig. \ref{fig:conv_gral} where we display the net flux as a function of the radius for different resolutions, 
we see the net flux outside the ergoregion manifest signs of convergence. 
At the BH horizon, on the other hand, the figure shows that there is a slight increase on the flux for higher resolutions.
\pagebreak

\subsection{Constraints preservation} \label{sec:constraints_preservation}

\begin{figure}
  \begin{center}
\includegraphics[scale=0.32]{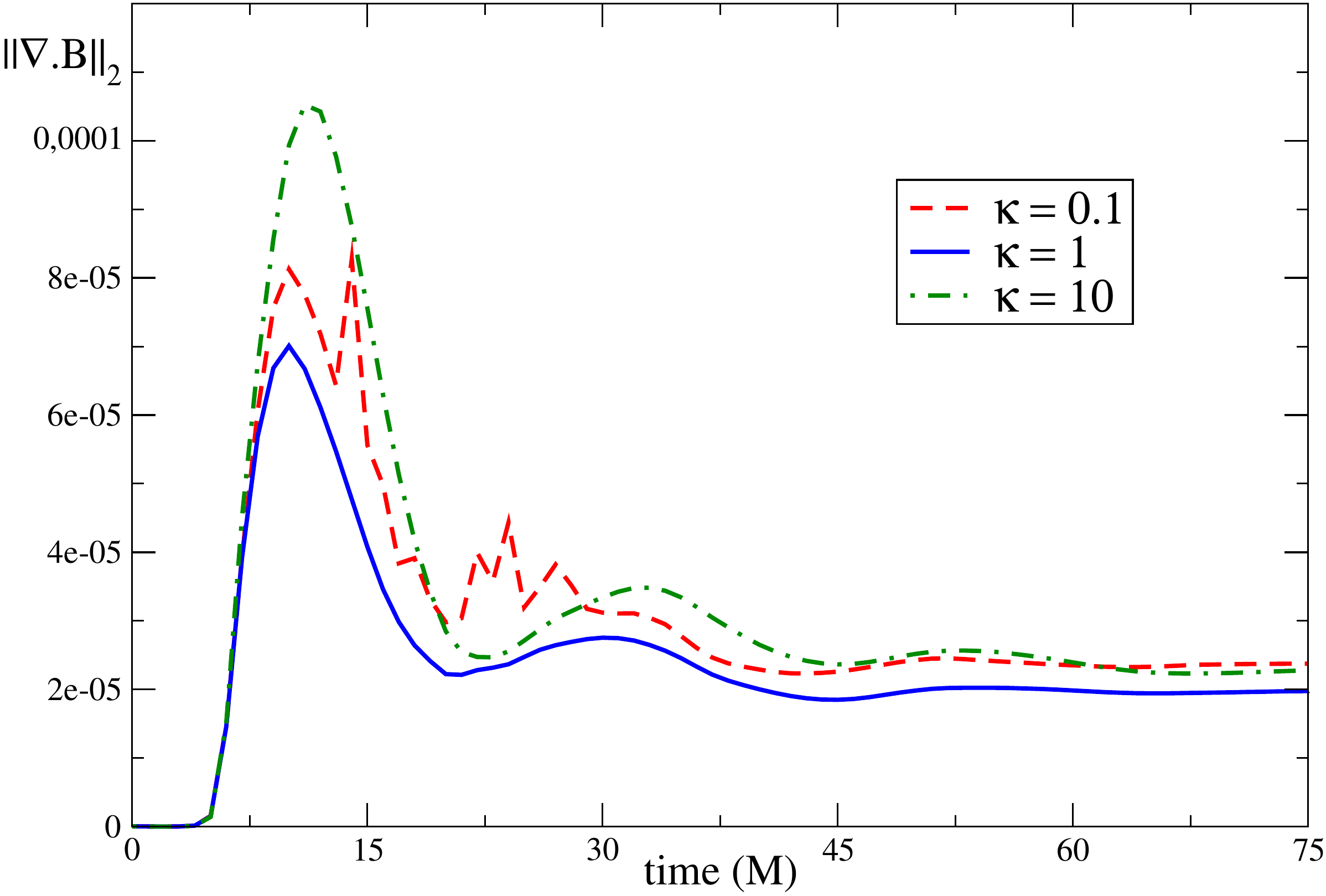}
\includegraphics[scale=0.32]{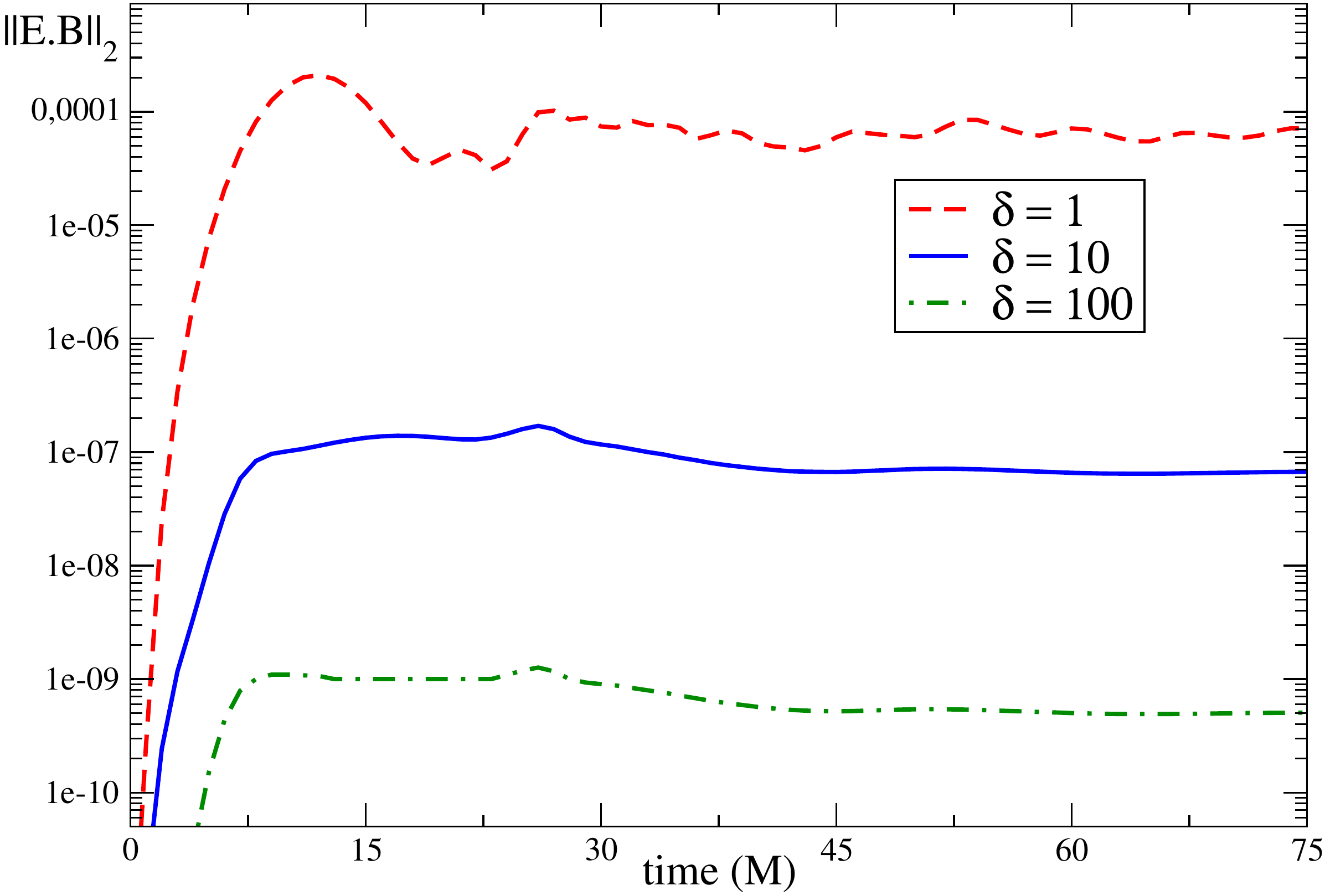}
  \caption{$L_2$ norms of the two constraints for a typical run. 
  \textbf{Top:} Evolution of $\frac{||\nabla\cdot\vec{B}||_{2}}{||B ||_{2}}$ for different values of $\kappa$. 
  \textbf{Bottom:} Evolution of $\frac{||\vec{E}\cdot \vec{B}||_2}{||B^2 ||_2}$ for different values of the parameter $\delta$.  }
 \label{fig:constraints} 
 \end{center}
\end{figure}

We monitor the two constraints of the theory (namely, $\nabla \cdot \vec{B} = 0$ and $G=0$) remain well behaved throughout the evolution.
On one hand, we check no significant constraint violations arise from the numerical boundary during the evolution:
modifications \eqref{normal_phi}-\eqref{normal_B} at boundary points has proved crucial in avoiding deviations of $\nabla \cdot \vec{B} = 0$ to access the domain.  
And on the other hand, we have verified no constraint violations develops within the whole integration region.
To see this, we have plotted, Fig.~\ref{fig:constraints}, the $L_2$ norm of relevant quantities associated with each of the constraints for a typical run. 
We have considered the quotient $(\nabla\cdot\vec{B})/|B|$, for different values $\kappa$, for the magnetic divergence-free constraint;
and $(\vec{E}\cdot \vec{B})/B^2$, for different values $\delta$ (see expression \eqref{damp}), as a measure of possible deviations of $G=0$. 

We find satisfactory results in both cases, showing a decent control on the constraints at the whole domain and through the evolution.
In particular, for the values of the parameter we have adopted in most of our simulations, namely: $\kappa=1$ and $\delta=100$.
Notice that even for small values of $\delta$ we still deviate little from $G=0$, since our evolution equations mathematically preserve this constraint 
and we have just included the extra term, equation \eqref{damp}, to have a better numerical control on it.


\bibliographystyle{unsrt} 
\bibliography{FFE}


\end{document}